\documentclass[
preprint,
unsortedaddress,
showkeys,
longbibliography,
amsmath,
amssymb,
pra,
floatfix,
]{revtex4-1}

\usepackage{graphicx}
\usepackage{dcolumn}
\usepackage{bm}
\usepackage{epsfig}
\usepackage{array}
\usepackage{mathrsfs}
\usepackage{amsmath}
\usepackage{verbatim}
\usepackage{longtable}
\usepackage{color}
\usepackage{soul}
\usepackage[normalem]{ulem}
\usepackage{natbib}

\newcommand{\stkout}[1]{\ifmmode\text{\sout{\ensuremath{#1}}}\else\sout{#1}\fi}

\DeclareFontFamily{OT1}{pzc}{}
\DeclareFontShape{OT1}{pzc}{m}{it}{<-> s * [1.10] pzcmi7t}{}
\DeclareMathAlphabet{\mathpzc}{OT1}{pzc}{m}{it}

\newcommand{\q}[2]
{
\frac{\partial #1}{\partial #2}
}
\newcommand{\dd}[2]
{
\frac{\rd#1}{\rd#2}
}
\newcommand{\p}[1]
{
p_k^{(#1)}
}

\newcommand{\aref}{a_{\rm ref}}
\newcommand{\thetaref}{\theta_{\rm ref}}
\newcommand{\tref}{t_{\rm ref}}
\newcommand{\uref}{u_{\rm ref}}
\newcommand{\cref}{c_{\rm ref}}
\newcommand{\cv}{c^{\rm (v)}}
\newcommand{\cvstar}{c^{{\rm (v)}*}}
\newcommand{\cvsat}{c^{\rm (v)}_{\rm sat}}
\newcommand{\cvinf}{c^{\rm (v)}_\infty}
\newcommand{\pa}{p_{\rm a}}
\newcommand{\Ca}{\mathop{\text{Ca}}\nolimits}
\newcommand{\Pe}{\mathop{\text{Pe}}\nolimits}
\newcommand{\xS}{x_{\rm S}}
\newcommand{\yS}{y_{\rm S}}
\newcommand{\Dv}{D^{\rm (v)}}
\newcommand{\Dp}{D^{\rm (p)}}
\newcommand\kth{k^{\rm th}}
\newcommand\nth{n^{\rm th}}
\newcommand\calJ{{\cal J}}
\newcommand\rd{\mathrm{d}}
\newcommand\ee{\mathrm{e}}
\newcommand\xihat{\hat{\xi}}
\newcommand\phiC{\phi_\text{C}}
\newcommand\phiS{\phi_\text{S}}
\newcommand\rS{r_\text{S}}

\newcommand\etal{{\it et al.\ }}
\newcommand\ie{{\it i.e.\ }}
\newcommand\eg{{\it e.g.\ }}

\begin{document}

\title{Contact-line deposits from multiple evaporating droplets}
\author{Alexander W.\ Wray}
\email{Email: alexander.wray@strath.ac.uk}
\affiliation{Department of Mathematics and Statistics,
University of Strathclyde,\\
Livingstone Tower,
26 Richmond Street,
Glasgow,
G1 1XH, UK}
\author{Patrick S.\ Wray}
\email{Email: wraypa1@gmail.com}
\affiliation{Drug Product Science and Technology,
Bristol-Myers Squibb,\\
Reeds Lane,
Moreton,
Wirral,
CH46 1QW, UK}
\author{Brian R.\ Duffy}
\email{Email: b.r.duffy@strath.ac.uk}
\author{Stephen K.\ Wilson}
\email{Author for Correspondence. Email: s.k.wilson@strath.ac.uk}
\affiliation{Department of Mathematics and Statistics,
University of Strathclyde,\\
Livingstone Tower,
26 Richmond Street,
Glasgow,
G1 1XH, UK}

\date{8th January 2021, revised 14th June 2021}

\begin{abstract}
Building on the recent theoretical work of
Wray \etal [J.\ Fluid Mech.\ \textbf{884}, A45 (2020)]
concerning the competitive diffusion-limited evaporation of
multiple thin sessile droplets in proximity to each other,
we obtain theoretical predictions for the spatially non-uniform
densities of the contact-line deposits
(often referred to as ``coffee stains'' or ``ring stains'')
left on the substrate after such droplets containing suspended
solid particles have completely evaporated.
Neighbouring droplets interact via their vapour fields,
which results in a spatially non-uniform ``shielding'' effect.
We give predictions for the deposits
from a pair of identical droplets, which show
that the deposit is reduced the most where the droplets are closest together, and
demonstrate excellent quantitative agreement with experimental results of
Pradhan and Panigrahi [Coll.\ Surf.\ A \textbf{482}, 562--567 (2015)].
We also give corresponding predictions for a triplet of identical
droplets arranged in an equilateral triangle, which show
that the effect of shielding on the deposit is more subtle in this case.
\end{abstract}


\maketitle

\clearpage

\section{Introduction}
\label{sec:introduction}

The evaporation of sessile droplets has been the subject of extensive
experimental, numerical and analytical investigation in recent years
(see, for example, \citep{
routh2013drying,
larson2014transport,
stauber2014lifetimes,
brutin2018recent,
giorgiutti2018drying},
and the references therein),
partly motivated by the wide range of everyday and
industrial situations, such as
protein crystallography \citep{dimitrov1994observations},
surface patterning \citep{boneberg1997formation},
ink-jet printing, including that of OLED displays \citep{bale2006ink}, and
agrochemical spraying of plants \citep{tredenick2020leaves},
in which it occurs.

Particular attention has been paid to the so-called
``coffee-stain'' or ``ring-stain'' effect: when a droplet of coffee
(or indeed a droplet of any fluid containing suspended solid particles)
with a pinned (\ie a fixed) contact line
evaporates it tends to deposit the majority of the particles close
to the location of its contact line, even if the particles were
initially distributed uniformly throughout the bulk of the droplet.
The explanation of this phenomenon,
as first given by \citet{deegan1997capillary},
is that as the droplet evaporates its free surface adjusts quasi-statically
under the effect of capillarity, inducing a flow within the droplet
that advects the particles suspended within it towards its contact line,
resulting in a characteristic ring-like contact-line deposit on the substrate after
the droplet has completely evaporated.
Since the seminal work by \citet{deegan1997capillary},
many aspects of this phenomenon have been investigated in considerable detail
(see, for example, \citep{
deegan2000patterns,
deegan2000CLdeposits,
popov2005evaporative,
zheng2009pipes,
askounis2011microscopy,
hamamoto2011velocity,
marin2011ordertodisorder,
marin2011rushhour,
yunker2011nature,
berteloot2012coffee,
askounis2013structural,
wray2014electrostatic,
boulogne2016coffee,
kang2016alternative,
kim2018coffee}
and the reviews \citep{
larson2014transport,
mampallil2018review,
yang2021review}).
Note that although \citet{deegan1997capillary} considered the
most commonly studied situation of diffusion-limited evaporation into
a quiescent atmosphere with a uniform far-field concentration of vapour,
which has a large (theoretically singular) evaporative flux
close to the contact line, the effect is quite robust, and even,
for example, a spatially uniform evaporative flux
will lead to advection of particles towards the contact line
(see, for example, \citep{
deegan2000CLdeposits,
zheng2009pipes,
boulogne2016coffee}).

The vast majority of the previous work on deposition from
evaporating sessile droplets has, for obvious reasons,
focused on axisymmetric deposits from axisymmetric droplets.
There has, however,
been some work on non-axisymmetric deposits from non-axisymmetric droplets
(see, for example, \citep{
deegan1997capillary,
cheng2008nanopatterning,
du2015ring,
kim2017inclined,
saenz2017dynamics,
timm2019evaporation,
tredenick2020leaves}).
In particular,
\citet{du2015ring}
examined a two-dimensional droplet on an inclined substrate numerically,
and found that, depending on the initial volume of the droplet and
the angle of inclination of the substrate, the larger deposit can
occur at either the upper or the lower contact line,
while
\citet{saenz2017dynamics}
investigated a variety of non-axisymmetric droplets both experimentally
and numerically, and found that larger deposits occur where the contact
line has the largest curvature
(\eg near the tips of a droplet with a triangular contact line).
However, their theoretical modelling of the density of the deposit
was essentially phenomenological.

Non-axisymmetric deposits also occur as a result of
the non-axisymmetric evaporation of multiple droplets in proximity to each other,
a situation that occurs much more commonly in practice than
single droplets in isolation \citep{bale2006ink}.
Specifically,
neighbouring droplets undergoing diffusion-limited evaporation interact
via their vapour fields, which results in a spatially non-uniform
``shielding'' effect that reduces the evaporation rate.
While there have been some analytical studies of mathematically analogous
situations concerning clusters of micro-contacts and nanobubbles
(see, for example, \citep{
argatov2011electrical,
dollet2016pinning}),
analytical work on the evaporation of multiple droplets is rather limited.
To a large extent this is explained by the inherent difficulty of analysing
such situations, and while the evaporation of multiple droplets in
various configurations has been the subject of growing recent interest
(see, for example, \citep{
lacasta1998competitive,
schafle1999cooperative,
deegan2000CLdeposits,
kokalj2010biologically,
sokuler2010dynamics,
pradhan2015deposition,
carrier2016collective,
castanet2016row,
shaikeea2016evaporating,
shaikeea2016insights,
hatte2019arrays,
khilifi2019study,
pandey2020arrays,
schofield2020shielding,
wray2020competitive}),
the previous studies have been predominantly numerical or experimental.
Two notable exceptions are the recent work of
\citet{wray2020competitive},
who, building on the earlier work of
\citet{fabrikant1985potential}
concerning a model for diffusion through a porous membrane,
analysed the spatially non-uniform shielding that occurs in
arbitrary configurations of thin droplets with circular contact lines,
and that of
\citet{schofield2020shielding},
who used conformal-mapping techniques
to analyse the analogous spatially non-uniform shielding that occurs in
the closely related two-dimensional situation of a pair of evaporating ridges.
In particular,
\citet{wray2020competitive}
gave explicit formulae for the evaporative flux of arbitrary
configurations of droplets that were found to be
remarkably accurate up to and including the limit of touching droplets,
and led to theoretical predictions for the evolution of
an arrangement of seven droplets that were found to be in excellent agreement
with experimental results of \citet{khilifi2019study}.

In the present contribution
we build on the work of \citet{wray2020competitive} in order
to analyse the spatially non-uniform densities of the deposits
left on the substrate by the diffusion-limited evaporation of
multiple thin droplets with pinned circular contact lines
in proximity to each other.
Specifically,
in Secs.~\ref{sec:formulation} and \ref{sec:pressure} we formulate
and solve the evaporation, hydrodynamic, and particle-transport problems.
In Sec.~\ref{sec:pair} we give theoretical predictions for the
densities of the deposits from a pair of identical droplets, and
demonstrate excellent quantitative agreement with experimental
results of \citet{pradhan2015deposition}.
In Sec.~\ref{sec:triplet}
we also give corresponding predictions for a triplet of identical droplets
arranged in an equilateral triangle.
Finally, we summarise our conclusions in Sec.~\ref{sec:conclusions}.

The present analysis is for
the most commonly studied case of small droplets,
in which capillary effects dominate over gravitational effects,
corresponding to the limit of small Bond number.
In Appendix \ref{sec:applargeBo} we describe
the corresponding analysis for
the less commonly studied case of large droplets,
corresponding to the limit of large Bond number,
in which even greater analytical progress is possible.

\section{Problem Formulation}
\label{sec:formulation}

\begin{figure}[tp]
\begin{center}
\begin{tabular}{cc}
\includegraphics[width=0.8\textwidth]{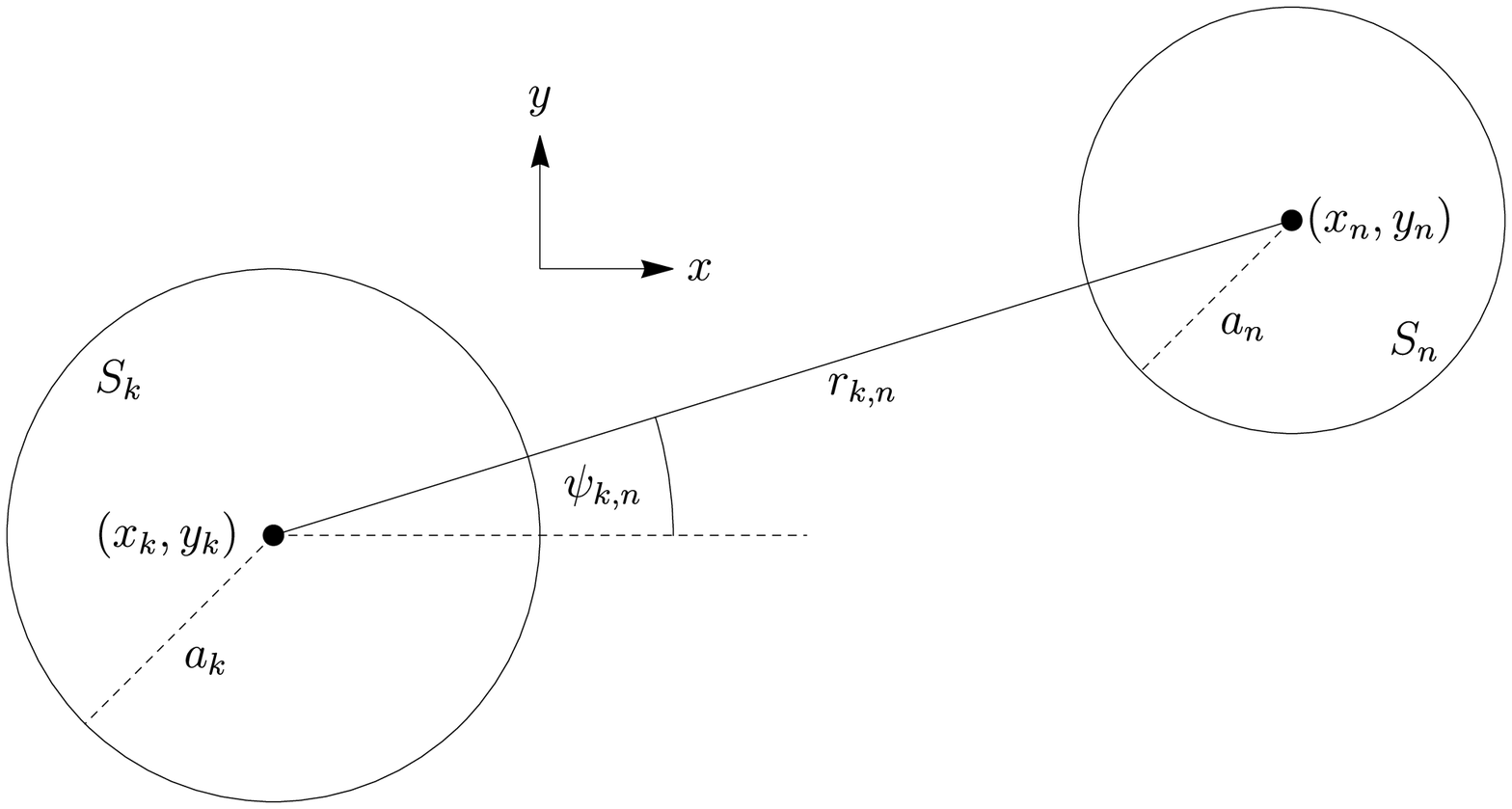}\\
(a)\\
\vspace{0.5cm}\\
\includegraphics[width=0.95\textwidth]{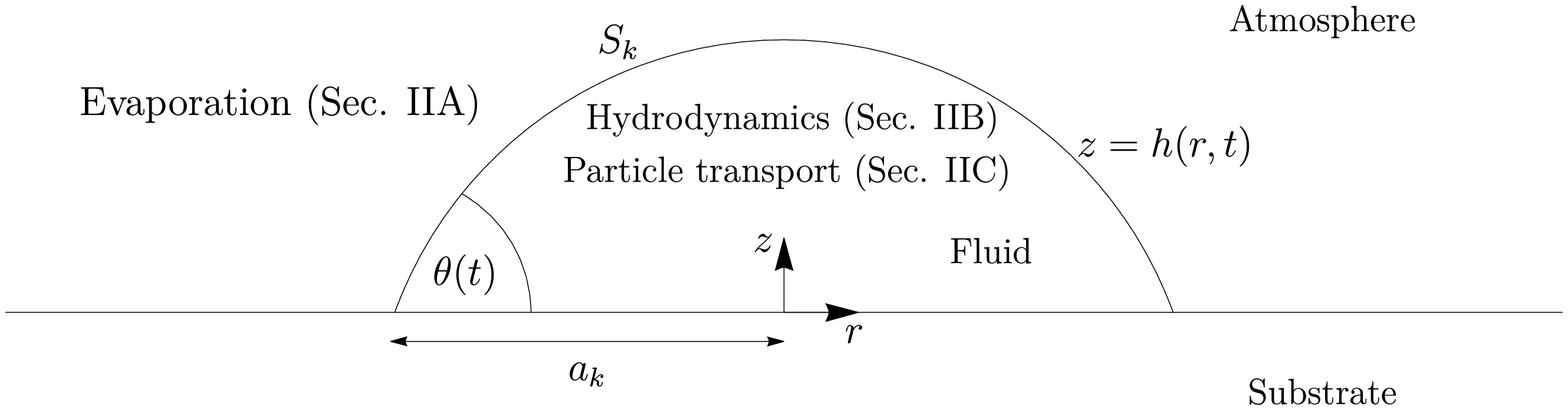}\\
(b)\\
\end{tabular}
\end{center}
\caption{
(a) Geometry of the $\kth$ and $\nth$ droplets on the substrate $z = 0$.
(b) A slice through the $\kth$ droplet showing the geometry of
the droplet and the problems to be solved in the different regions.
}
\label{fig:schematic}
\end{figure}

Consider $N$ ($N = 1, 2, 3, \ldots$) thin axisymmetric sessile droplets
with pinned circular contact lines with constant radii $a_k$
and fixed centres at $(x_k,y_k)$ for $k = 1, 2, \ldots, N$
on a planar solid substrate $z = 0$,
as shown in Fig.~\ref{fig:schematic}(a).
The droplets undergo quasi-static diffusion-limited evaporation,
which, as described in Sec.~\ref{sec:introduction},
induces flows within the droplets that advect the solid particles
suspended within them towards their contact lines.
The goal of the present work is to determine the
spatially non-uniform densities of the deposits left on
the substrate after the droplets have completely evaporated.
Three coupled problems must therefore be solved:
the evaporation problem for the concentration of vapour in the atmosphere
(which determines the rates of evaporation of the droplets),
the hydrodynamic problem for the fluid flow that is induced in each droplet, and
the advection problem for the motion of the particles suspended within each droplet,
as shown in Fig.~\ref{fig:schematic}(b).
We now discuss each of these problems in turn.

\subsection{The evaporation problem}
\label{sec:evaporation}

According to the diffusion-limited model,
the quasi-static concentration of vapour in the atmosphere,
denoted by $\cv = \cv(r,\phi,z)$,
satisfies Laplace's equation $\nabla^2\cv=0$ subject to
conditions of complete saturation at the free surfaces of the droplets and
of no flux of vapour through the unwetted part of the substrate.
Following \citet{wray2020competitive},
we scale and nondimensionalise variables appropriately for the atmosphere
according to
\begin{align}\label{eq:ScalingsEvap}
\begin{gathered}
x = \aref x^*, \quad
y = \aref y^*, \quad
z = \aref z^*, \quad
r = \aref r^*, \\
\cv = \cvinf + (\cvsat - \cvinf)\cvstar, \quad
J_k = \frac{\Dv(\cvsat - \cvinf)}{\aref}J_k^*, \quad
F_k = \Dv(\cvsat - \cvinf)\aref F_k^*,
\end{gathered}
\end{align}
where
$x$ and $y$ are Cartesian coordinates in the plane $z=0$,
$r$, $\phi$ and $z$ are local polar coordinates
with their origin at the centre of the $\kth$ droplet,
$\aref$ is a characteristic radius of the contact lines of the droplets,
$\Dv$ is the constant diffusion coefficient of vapour in the atmosphere,
$\cvsat$ and $\cvinf$ are
the constant saturation concentration and far-field concentration of vapour, and
$J_k = J_k(r,\phi)$ and $F_k$ are
the local evaporative flux and
the integral evaporative flux from the $\kth$ droplet, respectively,
which are related by
\begin{equation}\label{eq:Fk_definition}
F_k = \int\!\!\!\int_{S_k} J_k \, \rd{S},
\end{equation}
where $S_k$ denotes the free surface of the $\kth$ droplet.
Since the contact lines of the droplets are pinned,
$J_k$ and $F_k$ are independent of time except for discontinuous jumps when any droplet completely evaporates;
in particular, $J_k$ and $F_k$ jump instantaneously to zero when the $\kth$ droplet completely evaporates.

For clarity,
we immediately drop the star superscripts on non-dimensional quantities,
and so the boundary conditions on $\cv$ become
$\cv = 1$ on $z = h_k$ for $k = 1,2, \ldots, N$,
$\partial{\cv}/\partial{z} = 0$ on the unwetted part of the substrate $z=0$,
and
the far-field condition $\cv \to 0$ as $r^2+z^2 \to \infty$.

As \citet{wray2020competitive} described, the earlier work of
\citet{fabrikant1985potential} on diffusion through a porous membrane,
when interpreted in terms of the evaporation of multiple thin sessile droplets,
shows that the integral evaporative flux $F_k$ is given,
to a high degree of accuracy,
by the solution of the linear system
\begin{equation}
F_k = 4a_k - \frac{2}{\pi}\sum_{n=1, n\neq k}^N F_n\arcsin\left(\frac{a_k}{r_{k,n}}\right)
\quad \hbox{for} \quad k = 1, 2, \dots, N,
\end{equation}
where
$r_{k,n}$ ($\ge a_k + a_n$)
is the distance between the centres of the $\kth$ and the $\nth$ droplets
shown in Fig.~\ref{fig:schematic}(a) and given by
\begin{equation}
r_{k,n} = \sqrt{(x_n - x_k)^2 + (y_n - y_k)^2}.
\end{equation}
\citet{wray2020competitive} also showed that,
to the same high degree of accuracy,
the local evaporative flux $J_k$ is given by
\begin{align}\label{eq:sigmaSolAsymp}
J_k(r,\phi) &= \calJ_k(r) \left[1 - \sum_{n=1, n\neq k}^N
\frac{F_n\sqrt{r_{k,n}^2 - a_k^2}}
{2\pi\left(r^2 + r_{k,n}^2 - 2r r_{k,n}\cos(\phi - \psi_{k,n})\right)}\right],
\end{align}
where
\begin{equation}
\calJ_k(r) = \frac{2}{\pi\sqrt{a_k^2 - r^2}}
\end{equation}
is the local evaporative flux from the $\kth$ droplet in isolation, and
$\psi_{k,n}$ is the angle between the $x$ axis and
the line joining the centres of the $\kth$ and the $\nth$ droplets,
also shown in Fig.~\ref{fig:schematic}(a) and given by
\begin{equation}
\tan \psi_{k,n} = \frac{y_n - y_k}{x_n - x_k}.
\end{equation}

\subsection{The hydrodynamic problem}
\label{sec:hydrodynamics}

The velocity and pressure within the $\kth$ droplet,
denoted by
$\mathbf{u}_k = \mathbf{u}_k(r,\phi,z,t)
= u_k\mathbf{e}_r + v_k\mathbf{e}_\phi + w_k\mathbf{e}_z$
and
$p_k = p_k(r,\phi,z,t)$,
where $t$ denotes time,
satisfy the usual mass-conservation and Stokes equations
subject to the usual boundary conditions, and
the free surface, contact angle and volume of the $\kth$ droplet are denoted by
$z = h_k = h_k(r,t)$,
$\theta_k = \theta_k(t)$ ($\ll 1$) and
$V_k = V_k(t)$, respectively.

We scale and nondimensionalise variables appropriately for the droplet
according to
\begin{align}\label{eq:ScalingsHydro}
\begin{gathered}
z = \thetaref\aref \hat{z}, \quad
r = \aref \hat{r}, \quad
t = \tref \hat{t}, \quad
h_k = \thetaref\aref \hat{h}_k, \quad
V_k = \thetaref\aref^3 \hat{V}_k,
\\
u_k = \uref \hat{u}_k, \quad
v_k = \uref \hat{v}_k, \quad
w_k = \thetaref\uref \hat{w}_k, \quad
p_k - \pa = \frac{\gamma\thetaref}{\aref}\hat{p}_k,
\end{gathered}
\end{align}
in which $\thetaref$ ($\ll 1$) is
a characteristic contact angle of the droplets,
$\gamma$ is the constant coefficient of surface tension of the fluid,
$\pa$ is the constant atmospheric pressure, and
$\tref$ and $\uref$ are a characteristic time for the evaporation
and a characteristic velocity, defined by
\begin{equation}
\tref = \frac{\rho\thetaref\aref^2}{\Dv(\cvsat - \cvinf)},
\quad
\uref = \frac{\aref}{\tref} = \frac{\Dv(\cvsat - \cvinf)}{\rho\thetaref\aref},
\end{equation}
respectively,
where $\rho$ is the constant density of the fluid.

At leading order in $\thetaref \ll 1$
the governing equations for the $\kth$ droplet are,
with the hats dropped for clarity,
\begin{equation}\label{eq:vel_pdes}
\frac{1}{r}\q{(ru_k)}{r} + \frac{1}{r}\q{v_k}{\phi} + \q{w_k}{z} = 0, \quad
\Ca\q{^2u_k}{z^2} = \q{p_k}{r}, \quad
\Ca\q{^2v_k}{z^2} = \frac{1}{r}\q{p_k}{\phi}, \quad
\q{p_k}{z} = 0,
\end{equation}
where $\Ca$ is an appropriate capillary number, defined by
\begin{equation}\label{eq:Ca}
\Ca = \frac{\mu\uref}{\gamma\thetaref^3} =
\frac{\mu\Dv(\cvsat - \cvinf)}{\gamma\rho\thetaref^4\aref},
\end{equation}
where $\mu$ is the constant viscosity of the fluid.
Equation (\ref{eq:vel_pdes}) is to be solved subject to
zero velocity at the substrate,
\begin{equation}\label{eq:vel_bc_0}
u _k = 0, \quad v_k = 0, \quad w_k = 0 \quad \mbox{at} \quad z = 0,
\end{equation}
balances of normal and tangential stress
at the free surface of the droplet,
\begin{equation}\label{eq:vel_bc_h}
p_k = -\frac{1}{r}\frac{\partial}{\partial r}
\left(r\frac{\partial h_k}{\partial r}\right), \quad
\frac{\partial u_k}{\partial z} = 0, \quad
\frac{\partial v_k}{\partial z} = 0
\quad \mbox{at}\quad z = h_k,
\end{equation}
and the kinematic condition,
\begin{equation}\label{eq:kin_cond}
\q{h_k}{t} + \frac{1}{r}\q{}{r}\left(r Q^{(r)}_k\right) +
\frac{1}{r}\q{}{\phi}\left(Q^{(\phi)}_k\right) = -J_k
\quad {\rm at}\quad z = h_k,
\end{equation}
where $Q^{(r)}_k = Q^{(r)}_k(r, \phi, t)$ and
$Q^{(\phi)}_k = Q^{(\phi)}_k(r, \phi, t)$, defined by
\begin{equation}\label{eq:fluxExps1}
Q^{(r)}_k = \int_0^{h_k} u_k \, \rd{z},
\quad
Q^{(\phi)}_k = \int_0^{h_k} v_k \, \rd{z},
\end{equation}
are the local radial and azimuthal volume fluxes of fluid
within the droplet.

We consider the situation in which capillary effects are strong,
corresponding to small values of the capillary number $\Ca$, and so
we seek asymptotic solutions of the form
\begin{equation}\label{eq:Ca_expn}
u_k = u_{k0} + \Ca u_{k1} + O(\Ca^2), \quad
v_k = v_{k0} + \Ca v_{k1} + O(\Ca^2), \quad
p_k = p_{k0} + \Ca p_{k1} + O(\Ca^2)
\end{equation}
in the limit $\Ca \to 0$.
As we shall see, for the analysis of particle transport
presented in Sec.~\ref{subsec:transport},
we require only the leading order velocity components $u_{k0}$
and $v_{k0}$, which in turn require $p_{k0}$ and $p_{k1}$.

At leading order in $\Ca \ll 1$,
equations (\ref{eq:vel_pdes}) and (\ref{eq:vel_bc_h}) show that
the leading-order pressure is independent of $r$, $\phi$ and $z$, and is given by
$p_{k0} = p_{k0}(t) = 2\theta_k/a_k$, and
the leading-order free surface $z = h_k(r,t)$
takes the familiar paraboloidal form
\begin{equation}\label{eq:hk}
h_k = \frac{\theta_k(a_k^2 - r^2)}{2a_k}.
\end{equation}
The leading-order volume of the droplet is therefore given by
\begin{equation}\label{eq:Vk}
V_k = \int_{\phi=0}^{\phi=2\pi}\int_{r=0}^{r=a_k} r \, h_k(r,t)
\,\rd{r}\,\rd{\phi} = \frac{\pi \theta_k a_k^3}{4}.
\end{equation}

At first order in $\Ca \ll 1$,
equations (\ref{eq:vel_pdes}), (\ref{eq:vel_bc_h}) and (\ref{eq:vel_bc_0}) lead to
\begin{equation}
p_{k1} = p_{k1}(r,\phi,t), \quad
u_{k0} = \frac{{p_{k1}}_r}{2}\left(z^2 - 2h_kz\right),
\quad v_{k0} = \frac{{p_{k1}}_\phi}{2r}\left(z^2 - 2h_kz\right),
\end{equation}
and so the leading-order local fluid fluxes (\ref{eq:fluxExps1}) are
\begin{equation}\label{eq:fluxExps2}
Q^{(r)}_k = -\frac{h_k^3}{3}{p_{k1}}_r,
\quad
Q^{(\phi)}_k = -\frac{h_k^3}{3}\frac{{p_{k1}}_\phi}{r}.
\end{equation}
Dropping the subscript ``1'' on $p_{k1}$ henceforth for clarity,
the kinematic condition (\ref{eq:kin_cond}) therefore gives
\begin{equation}\label{eq:kinPartial}
\q{h_k}{t} + \frac{1}{r}\q{}{r}\left(-\frac{rh_k^3}{3}{p_{k}}_r\right) +
\frac{1}{r}\q{}{\phi}\left( -\frac{h_k^3}{3}\frac{{p_{k}}_\phi}{r} \right)
= -J_k.
\end{equation}
We may use this condition to obtain the differential equation
satisfied by $p_k$ by noting that
$h_k = h_k(r,t)$ is independent of the azimuthal coordinate $\phi$,
and that, by global mass conservation, the volume $V_k$ satisfies
\begin{equation}\label{eq:dVdt}
\dd{V_k}{t} = - \int_{\phi=0}^{\phi=2\pi}\int_{r=0}^{r=a_k} r \, J_k
\,\rd{r}\,\rd{\phi},
\end{equation}
so that, with (\ref{eq:hk}) and (\ref{eq:Vk}),
$\partial{h_k}/\partial t$
may be written as
\begin{equation}\label{eq:dhdtequation}
\q{h_k}{t} = \frac{\rd{\theta_k}}{\rd{t}}\frac{a_k^2 - r^2}{2a_k}
= \frac{4}{\pi a_k^3}\frac{\rd{V_k}}{\rd{t}}
\frac{a_k^2 - r^2}{2a_k} = -\frac{2(a_k^2 - r^2)}{\pi a_k^4}
\int_{\phi=0}^{\phi=2\pi}\int_{r=0}^{r=a_k} r \, J_k
\,\rd{r}\,\rd{\phi}.
\end{equation}
Finally, therefore, the kinematic condition (\ref{eq:kinPartial})
may be expressed in the form
\begin{equation}\label{eq:pkequation}
\frac{1}{r}\q{}{r}\left(\frac{rh_k^3}{3}{p_k}_r\right) +
\frac{h_k^3}{3r^2}{p_k}_{\phi\phi} = J_k -
\frac{2(a_k^2 - r^2)}{\pi a_k^4}\int_{\phi=0}^{\phi=2\pi}\int_{r=0}^{r=a_k} r \, J_k
\,\rd{r}\,\rd{\phi},
\end{equation}
which is a partial differential equation for $p_k$,
with all of the other quantities in (\ref{eq:pkequation}) being known.
Once $p_k$ is determined from (\ref{eq:pkequation}),
the local fluid fluxes $Q^{(r)}_k$ and $Q^{(\phi)}_k$
are given by (\ref{eq:fluxExps2}).

The depth-averaged radial and azimuthal velocities,
denoted by $\bar{u}_k=\bar{u}_k(r,\phi,t)$ and $\bar{v}_k=\bar{v}_k(r,\phi,t)$,
are defined by
\begin{equation}\label{eq:ubarandvbar}
\bar{u}_k = \frac{1}{h_k}\int_0^{h_k} u_k \, \rd{z}
= \frac{Q_k^{(r)}}{h_k},
\quad
\bar{v}_k = \frac{1}{h_k}\int_0^{h_k} v_k \, \rd{z}
= \frac{Q_k^{(\phi)}}{h_k},
\end{equation}
respectively.
For future reference,
note that the streamlines of the depth-averaged flow
are determined by solving
\begin{equation}\label{eq:streamlines}
\dd{r}{\phi} = \frac{r\bar{u}_k}{\bar{v}_k} =
\frac{rQ_k^{(r)}}{Q_k^{(\phi)}} = \frac{r^2{p_k}_r}{{p_k}_\phi}.
\end{equation}
Since $\bar{u}_k$ and $\bar{v}_k$ have the same functional dependence on $h$,
the $\theta_k$ has cancelled out of (\ref{eq:streamlines}), and so
the streamlines depend on time only via changes in the flux $J_k$.
This means that in certain situations the computation of the streamlines
is simplified somewhat by the fact (mentioned earlier)
that $J_k$ and hence $F_k$ are independent of $t$
except for discontinuous jumps when any droplet completely evaporates.
Specifically, if the droplets are arranged in such a way that
all of them completely evaporate at exactly the same time then
their streamlines remain unchanged throughout the evaporation.
In this case the determination of
the density of the deposit reduces to performing a single integral,
as will be described in Sec.~\ref{sec:pairdensityofdeposit} below.

\subsection{The particle-transport problem}
\label{subsec:transport}

The motion of the particles suspended within each droplet
is due to a combination of advection by the flow and diffusion,
and so the concentration of particles in the $\kth$ droplet,
denoted by $c_k = c_k(r,\phi,z,t)$,
satisfies the (scaled) advection--diffusion equation
\begin{equation}\label{eq:advDiff}
\thetaref^2\Pe\left[\q{c_k}{t} + u_k\q{c_k}{r}
+ \frac{v_k}{r}\q{c_k}{\phi} + w_k\q{c_k}{z}\right]
= \thetaref^2\left[\frac{1}{r}\q{}{r}\left(r\q{c_k}{r}\right)
+ \frac{1}{r^2}\q{^2c_k}{\phi^2}\right] + \q{^2c_k}{z^2},
\end{equation}
in which
$c_k$ has been nondimensionalised according to $c_k = \cref c_k^*$,
where $\cref$ is a characteristic concentration of particles,
$\Pe = \uref\aref/\Dp$ is an appropriate P\'eclet number, and
$\Dp$ is the constant diffusion coefficient for the particles in the fluid,
and where the star subscript has again been dropped for clarity.
Equation (\ref{eq:advDiff}) is subject to conditions of
no flux of particles through either the free surface of the droplet
or the substrate,
\begin{equation}
\mathbf{n}_k\cdot\nabla c_k = \Pe c_k J_k
\quad \mbox{at} \quad z = h_k,
\quad
\q{c_k}{z} = 0
\quad \mbox{at}\quad z = 0,
\end{equation}
where $\mathbf{n}_k$ denotes the outward unit normal to the
free surface of the $\kth$ droplet.

As is well known (see, for example,
\cite{wray2014electrostatic}),
when the P\'eclet number $\Pe$ is such that
$\thetaref^2 \ll \thetaref^2\Pe \ll 1$,
the leading-order concentration of particles, $c_k = c_k(r,\phi,t)$,
is independent of $z$ and satisfies
\begin{equation}\label{eq:AdvEqn}
\q{c_k}{t} + \bar{u}_k\q{c_k}{r} + \frac{\bar{v}_k}{r}\q{c_k}{\phi}
= \frac{c_k J_k}{h_k},
\end{equation}
where the depth-averaged radial and azimuthal velocities
$\bar{u}_k$ and $\bar{v}_k$ are given by (\ref{eq:ubarandvbar}).
Equation (\ref{eq:AdvEqn}) may be solved by the method of characteristics:
\begin{equation}
\dd{c_k}{t} = \frac{c_k J_k}{h_k}
\quad \text{on the characteristics} \quad
\dd{r}{t} = \bar{u}_k
\quad \hbox{and} \quad
\dd{\phi}{t} = \frac{\bar{v}_k}{r},
\end{equation}
subject to a prescribed initial condition $c_k = c_k(r,\phi,0)$ at
$t = 0$.
For simplicity, in all of the results presented below
we assume that the initial concentration of particles
takes the same uniform value in all of the droplets,
which we may, without loss of generality,
take to be unity.

\section{Solution for the pressure $p_k$}
\label{sec:pressure}

In general,
the expression for the evaporative flux $J_k$
given in (\ref{eq:sigmaSolAsymp}) is rather complicated, and
precludes solving (\ref{eq:pkequation}) for the pressure $p_k$ in closed form;
however, we may determine $p_k$ to arbitrary accuracy as follows.

For $r \le a_k \, (< r_{k,n})$ we expand $J_k$
given by (\ref{eq:sigmaSolAsymp}) as the convergent series
\begin{align}\label{eq:JExp}
J_k(r,\phi)& = \calJ_k(r)\left[1 - \sum_{n=1, n\neq k}^N
\frac{F_n}{2\pi r_{k,n}}
\left(1 + \frac{2r\cos\left(\phi - \psi_{k,n}\right)}{r_{k,n}}+
\cdots \right)\right],
\end{align}
which we may rearrange in the form of a truncated Fourier series
\begin{equation}
J_k(r,\phi) = \calJ_k(r)\left[ j_0(r) + \sum_{n=1, n\neq k}^N\sum_{m=1}^M
j_{n,m}(r)\cos\left[m\left(\phi - \psi_{k,n}\right)\right] \right]
\end{equation}
for a chosen number of modes $M$ ($M = 1, 2, 3, \ldots$),
with \textit{known} functions $j_0 = j_0(r)$ and $j_{n,m} = j_{n,m}(r)$.
To determine $p_k$ we decompose it into a corresponding form, namely
\begin{equation}\label{eq:pp_FS_form}
p_k = \p0 + \sum_{n=1, n\neq k}^N\sum_{m=1}^M \p{n,m}(r)
\cos\left[m\left(\phi - \psi_{k,n}\right)\right],
\end{equation}
substitution of which into (\ref{eq:pkequation}) leads to a
sequence of differential equations for $\p0$ and $\p{n,m}$,
\begin{equation}\label{eq:p0_eqn}
\frac{1}{r}\dd{}{r}\left(\frac{rh_k^3}{3}\dd{\p0}{r}\right) =
\calJ_k(r)j_0(r) - \frac{4F_k}{\pi\theta_k a_k^3}\frac{a_k^2 - r^2}{2a_k}
\end{equation}
and
\begin{equation}\label{eq:inhomog_pnm_eqn}
\frac{1}{r}\dd{}{r}\left(\frac{rh_k^3}{3}\dd{\p{n,m}}{r}\right)
- \frac{m^2h_k^3}{3r^2}\p{n,m} = \calJ_k(r)j_{n,m}(r).
\end{equation}
Equation (\ref{eq:p0_eqn}) may be solved directly
(up to an irrelevant additive constant)
subject to regularity at the origin.
Equation (\ref{eq:inhomog_pnm_eqn}) may be solved
by the method of variation of parameters.
Specifically, the homogeneous version of
(\ref{eq:inhomog_pnm_eqn}), namely
\begin{equation}\label{eq:homog_p_eqn}
\dd{^2\p{n,m}}{r^2} + \left(\frac{1}{r} - \frac{6 r}{a_k^2 - r^2} \right)
\dd{\p{n,m}}{r} - \frac{m^2}{r^2}\p{n,m} = 0,
\end{equation}
has solutions $\p{n,m} = P_{k1}^{(m)}(r)$ and $\p{n,m} = P_{k2}^{(m)}(r)$
given by
\begin{equation}
P_{k1}^{(m)} =
    \left(\frac{r}{a_k}\right)^m
    \,_2F_1\left(
    \frac{1}{2}(3+M_+),
    \frac{1}{2}(3+M_-);
    m+1;
    \frac{r^2}{a_k^2}
    \right),
\end{equation}
which satisfies
\begin{equation}
P_{k1}^{(m)} \sim \left\{
    \begin{array}{ll}
    \displaystyle
    \left(\frac{r}{a_k}\right)^m
    & \quad \hbox{for}\displaystyle \quad \frac{r}{a_k} \ll 1, \\
    \displaystyle
    \frac{
    \Gamma(1+m)
    }{
    4\Gamma\left(\frac{1}{2}(3+M_-)\right)
     \Gamma\left(\frac{1}{2}(3+M_+)\right)
    }
    \left(\frac{a_k}{a_k-r}\right)^2
    & \quad \hbox{for}\displaystyle \quad 1-\frac{r}{a_k} \ll 1,
    \end{array}
\right.
\end{equation}
and
\begin{equation}
P_{k2}^{(m)} =
    \left(\frac{a_k}{r}\right)^m
    \,_2F_1\left(
    \frac{1}{2}(3-M_+),
    \frac{1}{2}(3-M_-);
    3;
    1-\frac{r^2}{a_k^2}
    \right),
\end{equation}
which satisfies
\begin{equation}
P_{k2}^{(m)} \sim \left\{
    \begin{array}{ll}
    \displaystyle
    \frac{2\Gamma(m)}{
    \Gamma\left(\frac{1}{2}(3+M_-)\right)
    \Gamma\left(\frac{1}{2}(3+M_+)\right)
    }
    \left(\frac{a_k}{r}\right)^m
    & \quad \hbox{for}\displaystyle \quad \frac{r}{a_k} \ll 1, \\
    1
    & \quad \hbox{for}\displaystyle \quad 1-\frac{r}{a_k} \ll 1,
    \end{array}
\right.
\end{equation}
where $M_{\pm} = m\pm\sqrt{m^2 + 9}$, and so the solution
of the inhomogeneous equation (\ref{eq:inhomog_pnm_eqn}) is
\begin{equation}\label{eq:inhomSol}
\p{n,m}(r) = 3P_{k2}^{(m)}(r) \int_0^r
\frac{\calJ_k(\tilde{r})j_{n,m}(\tilde{r})P_{k1}^{(m)}(\tilde{r})}
{W^{(m)}(\tilde{r}) h_k(\tilde{r})^3} \,\rd{\tilde{r}}
+ 3P_{k1}^{(m)}(r) \int_{r}^{a_k}
\frac{\calJ_k(\tilde{r})j_{n,m}(\tilde{r}) P_{k2}^{(m)}(\tilde{r})}
{W^{(m)}(\tilde{r}) h_k(\tilde{r})^3} \, \rd{\tilde{r}},
\end{equation}
where
\begin{equation}
W^{(m)} = P_{k1}^{(m)}\, \frac{\rd{P_{k2}^{(m)}}}{\rd{r}}
- P_{k2}^{(m)}\,\frac{\rd{P_{k1}^{(m)}}}{\rd{r}}
\end{equation}
is the Wronskian, which can be evaluated to give
\begin{equation}
W^{(m)} = - \frac{4\Gamma(m+1)a_k^6}
{\Gamma\left(\frac{1}{2}\left(3+M_+\right)\right)\Gamma\left(\frac{1}{2}\left(3+M_-\right)\right)r(a_k^2-r^2)^3}.
\end{equation}
Note that we have chosen the forms of the homogeneous solutions
and imposed the boundary conditions by selecting the lower limits
of the integrals in equation (\ref{eq:inhomSol}) so as to ensure
regularity at the origin (first integral) and at the contact line
(second integral).
In general, the integrals in equation (\ref{eq:inhomSol})
must be evaluated numerically to obtain $\p{n,m}(r)$.
Note, however, that,
as described in Appendix \ref{sec:applargeBo},
even greater analytical progress is possible
for the corresponding problem in the limit of large Bond number.

\section{A pair of identical droplets}
\label{sec:pair}

In this Section we apply the general methodology developed in
Secs.~\ref{sec:formulation} and \ref{sec:pressure} to determine
the densities of the deposits from a pair of identical droplets,
a situation for which the predictions of the present asymptotic theory
for the local evaporative flux $J_k$ and the integral evaporative
flux $F_k$ were validated by \citet{wray2020competitive}.
In Sec.~\ref{sec:pairfluxesandstreamlines} we determine
the local evaporative fluxes,
the fluid fluxes and
the resulting streamlines of the depth-averaged flows,
in Sec.~\ref{sec:pairdensityofdeposit} we determine
the density of the deposit, while
in Sec.~\ref{sec:paircomparision} we compare
the theoretical predictions for the density of the deposit with
the experimental results of \citet{pradhan2015deposition}.

\subsection{Evaporative fluxes, fluid fluxes and streamlines}
\label{sec:pairfluxesandstreamlines}

\begin{figure}[tp]
\begin{center}
\includegraphics[width=0.85\textwidth]{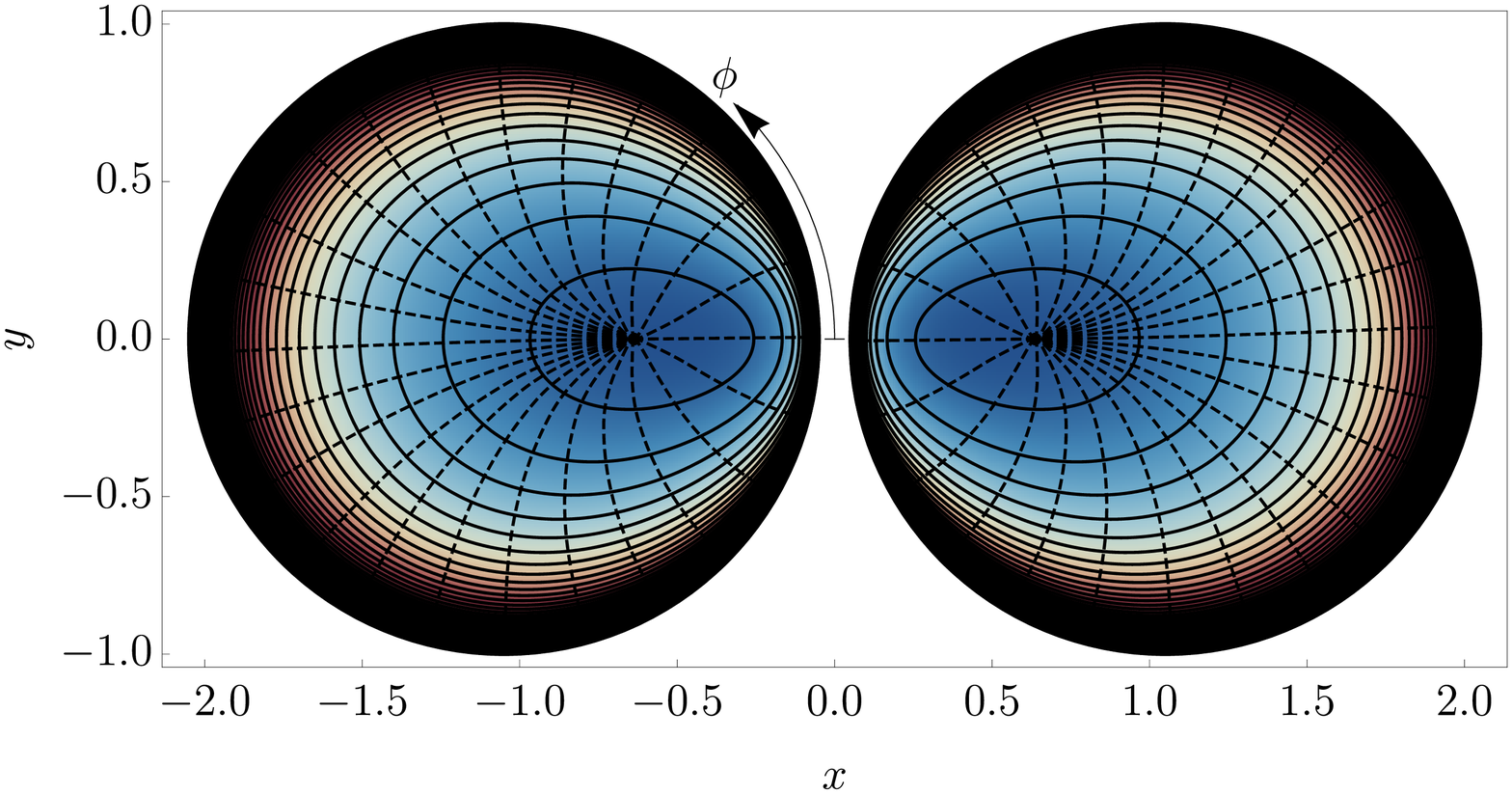} \hspace{0.25cm}
\raisebox{0.25\height}{\includegraphics[width=0.065\textwidth]{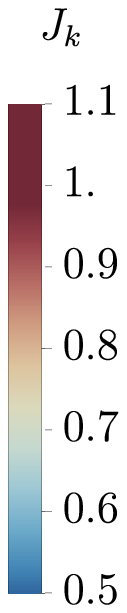}}
\end{center}
\caption{
Contours of the local evaporative fluxes
$J_1$ (left-hand droplet) and $J_2$ (right-hand droplet) (solid curves), and
the resulting streamlines of the depth-averaged flows (dashed curves)
for a pair of identical droplets of unit radius
with their centres a distance $b=2.1$ apart.
}
\label{fig:pairfluxesandstreamlines}
\end{figure}

Consider the evaporation of a pair of identical droplets,
which we may, without loss of generality,
take to be of unit radius $a_1=a_2=1$,
with their centres located at $(\pm b/2,0)$,
\ie with their centres a distance $r_{1,2}=r_{2,1}=b \, (>2)$ apart.
As \citet{wray2020competitive} showed,
the two droplets have the same integral evaporative flux
given by $F_1=F_2=F$, where
\begin{equation}\label{eq:integralTwoDropFlux}
F = \frac{4}{1 + (2/\pi)\arcsin(1/b)}.
\end{equation}
By symmetry,
it is sufficient to consider only the left-hand droplet
with its centre located at $(-b/2,0)$,
corresponding to $k=1$.
The local evaporative flux from the surface of the droplet is given by
\begin{equation}\label{eq:pairJ1}
J_1 = \calJ(r)\left[1 - \frac{F\sqrt{b^2 - 1}}
{2\pi(r^2 + b^2 - 2rb\cos\phi)}\right],
\end{equation}
which may be expanded as
\begin{equation}\label{eq:pairJ1expansion}
J_1 = \calJ(r)\sum_{m=0}^{M} j_m(r)\cos{m\phi},
\end{equation}
where
\begin{equation}\label{eq:j}
j_0 = 1 + \frac{F(1 - 2b^2 - 2r^2)}{4\pi b^3},
\quad
j_1 = \frac{F\left(1 - 2b^2 - 2r^2\right)r}{2\pi b^4},
\quad
j_2 = - \frac{F r^2}{\pi b^3},
\quad
j_3 = -\frac{F r^3}{\pi b^4},
\quad
\ldots,
\end{equation}
and
\begin{equation}\label{eq:isolatedJ}
\calJ(r) = \frac{2}{\pi\sqrt{1-r^2}}
\end{equation}
is the flux from the same droplet in isolation.
The expansion (\ref{eq:pairJ1expansion}) with (\ref{eq:j}) can then be used
to determine the pressure $p_1$ in (\ref{eq:pp_FS_form}),
and thus
the local fluid fluxes $Q^{(r)}_1$ and $Q^{(\phi)}_1$ in (\ref{eq:fluxExps2}).
Figure~\ref{fig:pairfluxesandstreamlines} shows
contours of the local evaporative fluxes
$J_1$ (left-hand droplet) and $J_2$ (right-hand droplet), and
the resulting streamlines of the depth-averaged flows
for $b = 2.1$,
with the expansion (\ref{eq:pairJ1expansion})
truncated after $M=10$ Fourier modes.
Figure~\ref{fig:pairfluxesandstreamlines} illustrates how
the shielding effect reduces the local evaporative flux
the most where the droplets are closest together
(\ie at $\phi=0$) and
the least where they are furthest apart
(\ie at $\phi=\pi$)
\cite{wray2020competitive}.
For an isolated droplet
the contours of $\calJ$ given by (\ref{eq:isolatedJ})
are concentric circles and the streamlines are radial lines;
Fig.~\ref{fig:pairfluxesandstreamlines} also illustrates how
the shielding effect skews both of them towards the other droplet.

\subsection{Density of the deposit}
\label{sec:pairdensityofdeposit}

\begin{figure}[tp]
\begin{center}
(a)
\includegraphics[width=0.9\textwidth]{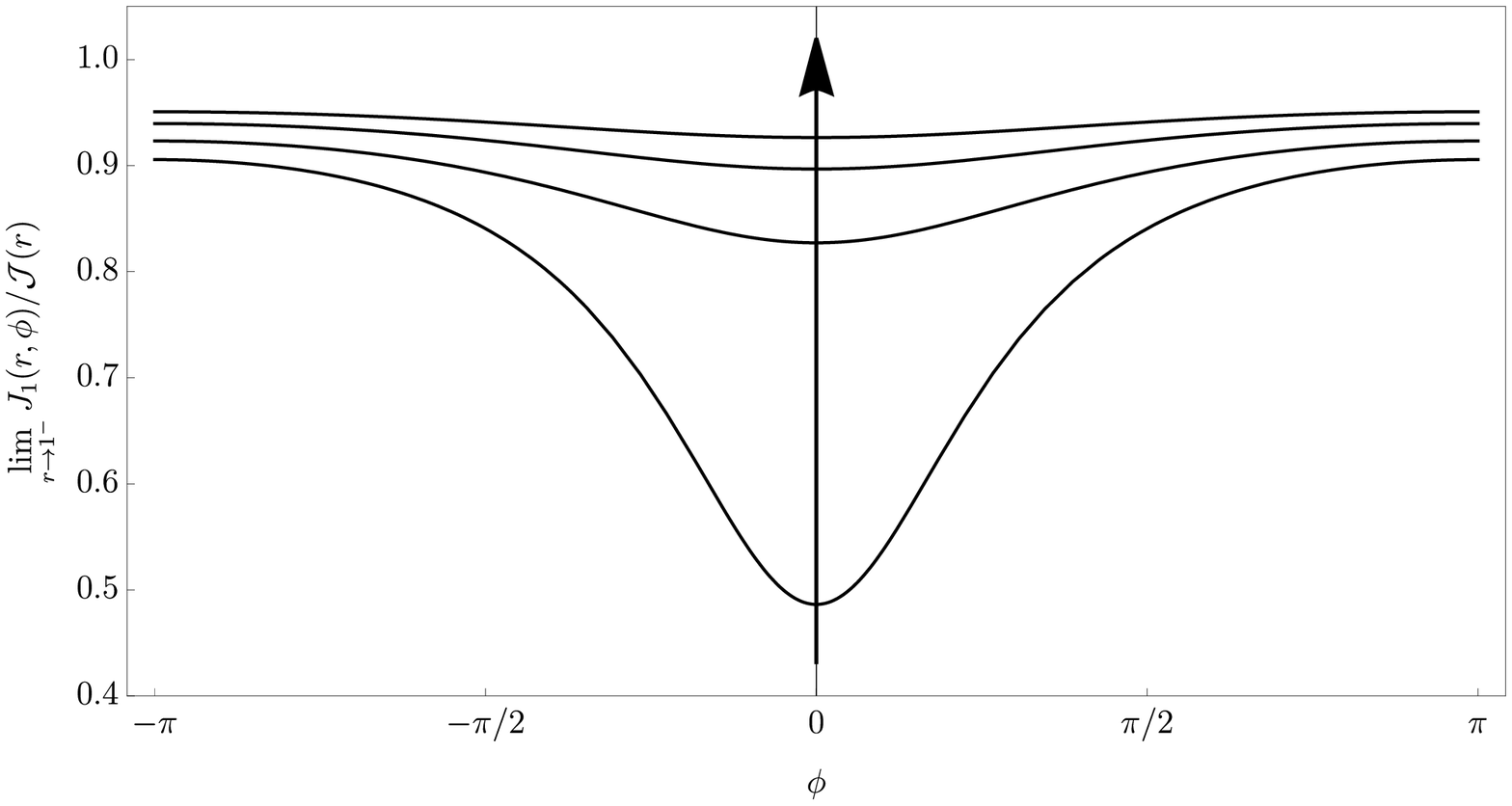}\\
(b)
\includegraphics[width=0.9\textwidth]{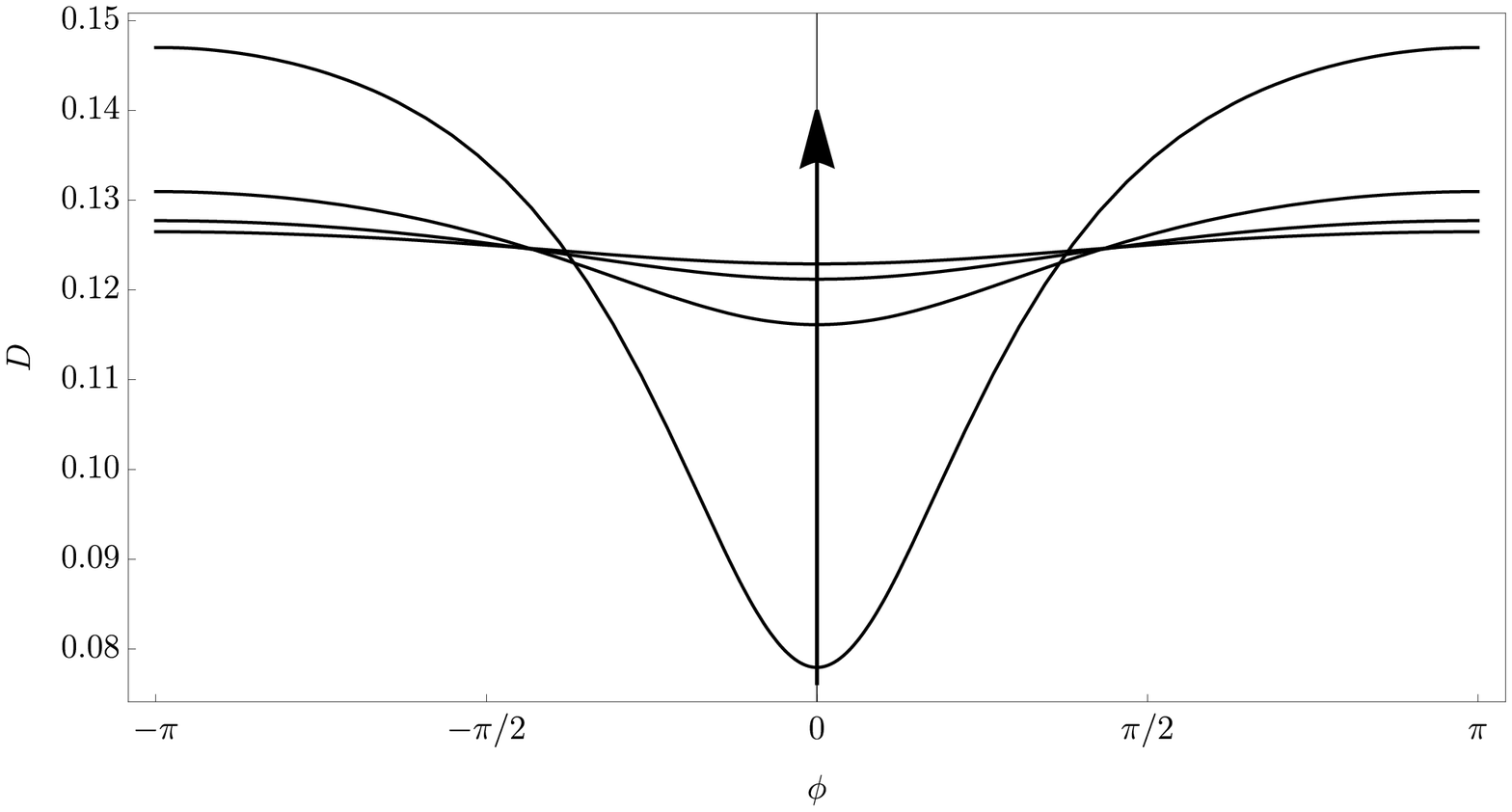}\\
\end{center}
\caption{
(a) The normalised evaporative flux at the contact line,
$\lim_{r \to 1^-}J_1(r,\phi)/\calJ(r)$,
and
(b) the density of the deposit $D$
as functions of the azimuthal coordinate $\phi$
for a pair of identical droplets of unit radius
with their centres distances $b = 2.5$, $5$, $7.5$, $10$ apart.
In both parts the arrow indicates the direction of increasing $b$.
}
\label{fig:pairfluxanddensityofdeposit}
\end{figure}

We now determine the density of the deposit
resulting from the evaporation of the droplets.
By symmetry,
the two droplets completely evaporate at the same time, and so
the streamlines of their depth-averaged flows remain unchanged throughout the evaporation
and, as in Sec.~\ref{sec:pairfluxesandstreamlines},
it is sufficient to consider only the left-hand droplet
with its centre located at $(-b/2,0)$.

We work relative to (non-orthogonal) curvilinear co-ordinates $(\chi,\xi)$
defined by the streamlines with their origin at the ``source''
from which all of the streamlines of the depth-averaged flow emanate, denoted by $(\xS,\yS)$,
shown in Fig.~\ref{fig:pairfluxesandstreamlines}.
Note that, contrary to what Fig.~\ref{fig:pairfluxesandstreamlines}
may suggest, the source from which the streamlines emanate, $(\xS,\yS)$,
does not, in general, coincide exactly with the location of the minimum
of the local evaporative flux. The difference between the two is readily
evident in Fig.~\ref{fig:tripletfluxesandstreamlines}
which appears subsequently in Sec.~\ref{sec:triplet}.
The coordinate $\xi \in [0,2\pi)$ parameterises the streamlines
such that the local behaviour of the streamlines near the source is given by
$(x,y) = (\xS,\yS) + \delta\left(\cos\xi,\sin\xi\right) + O(\delta^2)$
as $\delta \to 0^+$,
while
the coordinate $\chi \, (\ge 0)$ is the arc length along each streamline,
measuring from $\chi=0$ at the source to $\chi=\chi_\text{max}(\xi)$ at the contact line.
Note that $\chi_\text{max}(\xi)$ is therefore
the length of the streamline parameterised by $\xi$ from the source to the contact line.

Since the fluid flow always advects the particles
along the streamlines towards the contact line,
the mass of the deposit that eventually accumulates at the contact line between
the point with coordinate $\xi=0$ and
a general point with coordinate $\xi$,
denoted by $M=M(\xi)$,
is exactly equal to the mass of particles originally in the curved sector
between the streamlines parameterised by $\xi=0$ and by $\xi$, \ie
\begin{equation}\label{eq:M}
M(\xi) = \int_{\xihat=0}^{\xihat=\xi}\int_{\chi=0}^{\chi=\chi_\text{max}(\xihat)}
h\big[x(\chi,\xihat),y(\chi,\xihat)\big]\,
\frac{\partial(x,y)}{\partial(\chi,\xihat)}
\,\rd{\chi} \, \rd{\xihat}.
\end{equation}
Note that, by definition, $M(0)=0$ and $M(2\pi)=\pi/4$.
Once $M$ has been determined,
the density of the deposit
at a point on the contact line with polar angle $\phi$,
denoted by $D=D(\phi)$,
is then given by
\begin{equation}\label{eq:D}
D = \left.\frac{1}{r}\dd{M}{\phi}\right\vert_{r=1}
= \left.\frac{1}{r}\dd{M}{\xi}\q{\xi}{\phi}\right\vert_{r=1}.
\end{equation}

Figure~\ref{fig:pairfluxanddensityofdeposit} shows
the evaporative flux at the contact line
normalised by the corresponding flux for the same droplet in isolation,
$\lim_{r \to 1^-}J_1(r,\phi)/\calJ(r)$,
and
the density of the deposit $D$
as functions of the azimuthal coordinate $\phi$
for several values of $b$.
In particular,
Fig.~\ref{fig:pairfluxanddensityofdeposit}(b) shows that
the shielding effect described in Sec.~\ref{sec:pairfluxesandstreamlines}
and shown in Fig.~\ref{fig:pairfluxesandstreamlines}
leads to a spatially non-uniform deposit with
the smallest density where the shielding effect is strongest (\ie at $\phi=0$) and
the largest density where it is weakest (\ie at $\phi=\pi$).
Note that,
by conservation of mass, the total mass of the deposit
is the same for all of the values of $b$ used in Fig.~\ref{fig:pairfluxanddensityofdeposit}(b).

\subsection{Comparison with experimental results}
\label{sec:paircomparision}

\begin{figure}[tp]
\begin{center}
\includegraphics[width=0.95\textwidth]{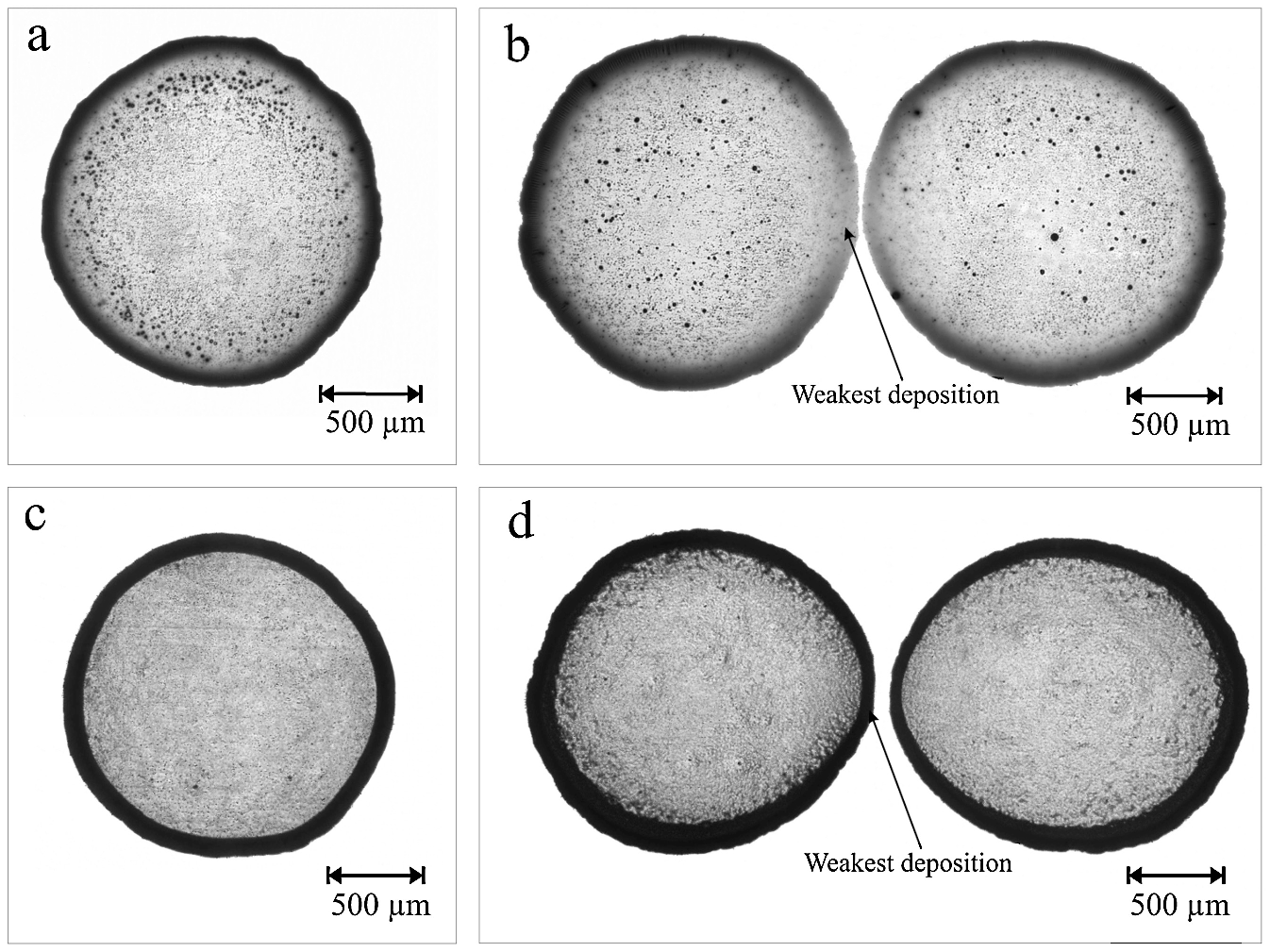}
\end{center}
\caption{
Figure 2 of \citet{pradhan2015deposition}, showing contact-line deposits from
(a) a single droplet and (b) a pair of droplets of ink with a minimum separation of 25 $\mu$m,
(c) a single droplet and (d) a pair of droplets of water containing 1 $\mu$m particles with a minimum separation of 95 $\mu$m.
Reprinted from {\it Colloids and Surfaces A: Physicochemical and Engineering Aspects}, Volume {\bf 482},
``Deposition pattern of interacting droplets'', Pages 562--567, Copyright 2015, with permission from Elsevier.
}
\label{fig:pradhan2015}
\end{figure}

In general,
comparing theoretical predictions for the density of a deposit with
experimentally obtained images of deposition patterns is a challenging task.
In particular,
because the depth of the deposit cannot usually be readily
determined from images taken from above the droplet, obtaining
a quantitative measure of the amount of deposit from experimental
images may often not be possible due to saturation and non-linearity of the data.
For example,
in Fig.~2 of \citet{pradhan2015deposition},
reproduced here as Fig.~\ref{fig:pradhan2015},
the image of the final deposition pattern from
a pair of similar droplets of water containing 1 $\mu$m particles
shown in part (d) has zero transmittance through the contact-line deposit,
making it impossible to quantify its spatial distribution.
However,
the image of the final deposition pattern from
a pair of similar droplets of ink
shown in part (b)
has nonzero transmittance throughout the vast majority of the contact-line deposit, and so
offers us an opportunity to quantify the relative amount of deposit as a function of azimuthal position.

In order to compare the results of \citet{pradhan2015deposition}
with the present theoretical predictions, it is first
necessary to convert the information contained within the
image into a form proportional to the density distribution $D$.
The raw data extracted from the image corresponds to
the {\it reflectance} of the deposit;
this must be converted into its {\it absorbance},
which can then be related to its concentration
via the Beer--Lambert law \citep{swinehart1962beer},
\begin{equation}\label{eq:BeerLambertLaw}
A = -\log_{10}\left(\frac{P}{P_0}\right) = \epsilon\,l\,C,
\end{equation}
where
$A$ is the absorbance,
$P$ is the radiant power (reflectance) or intensity of the light
as measured at each pixel in the image,
$P_0$ is the initial radiant power of the light
before absorbance,
$\epsilon$ is the molar absorptivity of the deposit
(which may reasonably be assumed to be constant),
$l$ is the path length of the light in the deposit, and
$C$ is the concentration of the absorbing species.
The values for $P$ are known, as this is the light gathered by
each pixel in the sensor of the camera to generate the image;
however, the value of $P_0$ is unknown.
(Ideally $P_0$ would have been determined through the
collection of a reference image in which the light reflected
from a calibrated sample, such as a 99\% reflectance standard,
was captured.)
Therefore, it is necessary to make an estimate of $P_0$, and
to do this we used the maximum possible brightness value of $255$ as
the reference value for all of the pixels (consistent with
the brightness of the image outside the footprint of the droplets).
In order to gather only density data from the deposit
near the contact line of each droplet,
and not the density of any residual deposit left within it,
the data was taken from an annular region
around the edge of the footprint of each droplet.
Slightly unfortunately,
as Fig.~\ref{fig:pradhan2015} shows,
in the published image the deposit from the left-hand droplet
is overlaid by the head of an arrow that the authors added to
indicate the region of weakest deposition.
In order to reduce artifacts associated with this arrowhead,
it was removed from the data by interpolating from
the neighbouring pixels.
The procedure for the extraction of the data for each droplet
was implemented in Python \cite{rossum1995python},
and is detailed below:
\begin{enumerate}
\item Convert the image to binary, and determine its centre of mass.
\item Trace out the outer perimeter of the deposit.
\item Define a second inner perimeter at $80\%$ of the radius of
the outer perimeter measured relative to the centre of mass.
\item Divide the annular region between the inner and outer perimeters
into $N$ sectors subtending equal angles at the centre of mass.
In practice, $N = 300$ sectors were used.
\item Use the Beer–Lambert law (\ref{eq:BeerLambertLaw}) to
calculate the absorbance of each pixel, which is proportional
to the mass of residue per unit area.
\item Integrate the mass per unit area numerically over each sector,
and divide by the angle subtended, to determine the corresponding
proportional mass per unit length of the contact line
(\ie the density to be compared with $D$).
\end{enumerate}

\begin{figure}[tp]
\begin{center}
\begin{tabular}{c}
(a)
\includegraphics[width=0.95\textwidth]{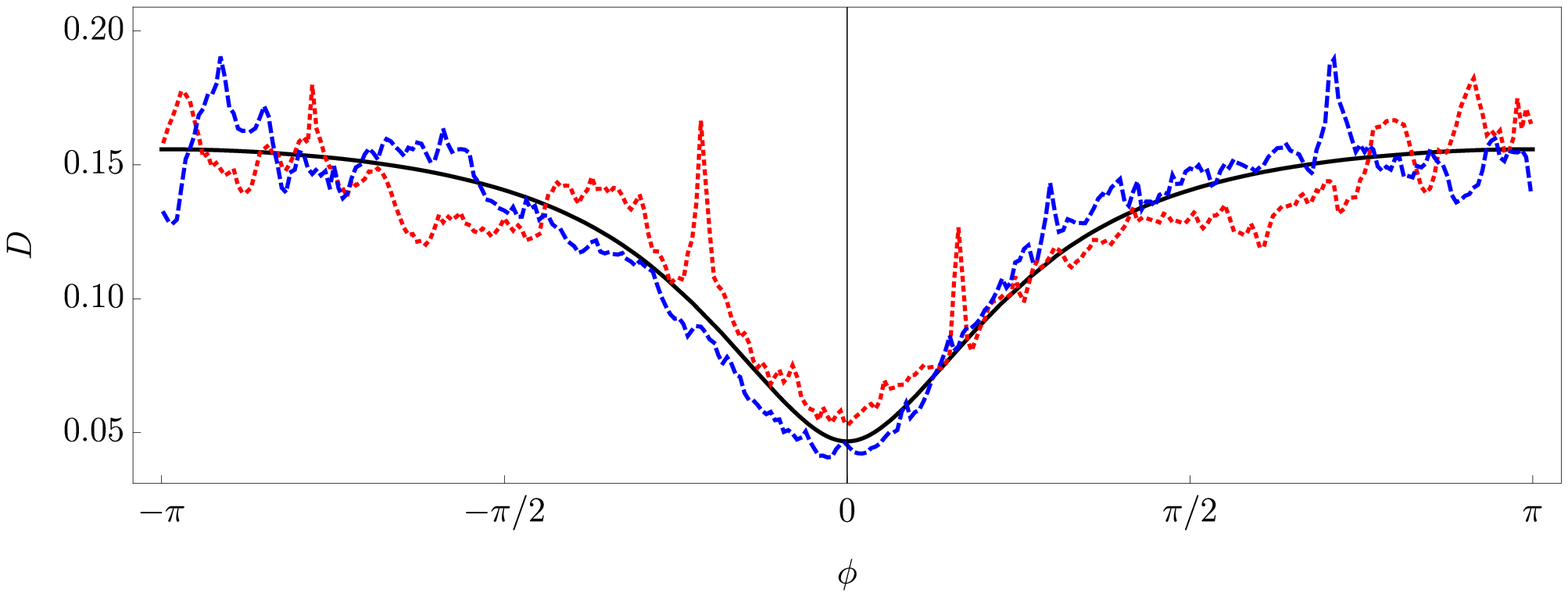}\\
\vspace{0.4cm}\\
(b)
\includegraphics[width=0.95\textwidth]{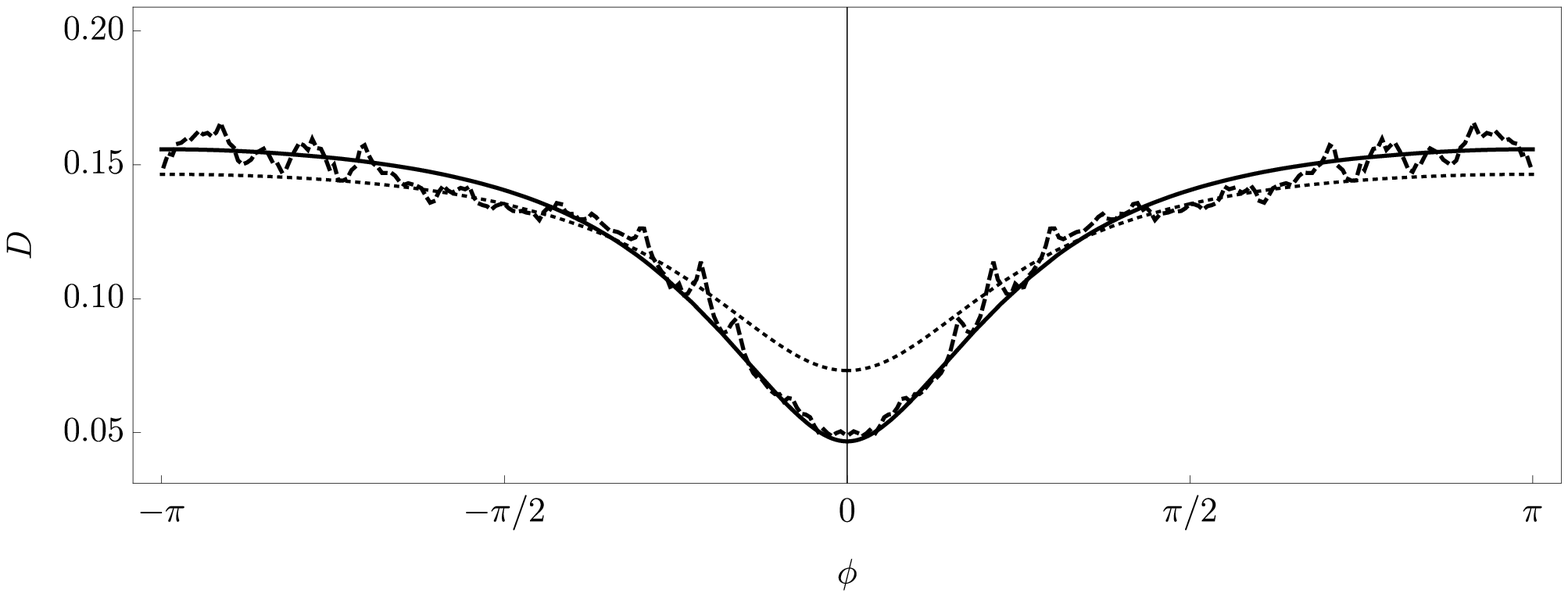}\\
\end{tabular}
\end{center}
\caption{
(a) Comparison between the density of the deposit
predicted by (\ref{eq:D}) (black solid curve) and
the experimental results for
the left-hand droplet (blue dashed curve) and
the right-hand droplet (red dotted curve)
extracted from Fig.~2(b) of \citet{pradhan2015deposition}.
(b) Comparison between the density of the deposit
predicted by (\ref{eq:D}) (black solid curve) and
the averaged experimental results for
both droplets shown in part (a) (black dashed curve), and
the corresponding phenomenological estimate
derived by \citet{wray2020competitive}
given by (\ref{eq:estimate}) (black dotted curve).
}
\label{fig:comp}
\end{figure}

The constant of proportionality for comparison with $D$
is determined by imposing the condition that
the integral of the density is $\pi/4$, in line with
the nondimensionalisation in Sec.~\ref{subsec:transport}.
Figure~\ref{fig:comp} shows a comparison between
the present theoretical predictions and
the experimental results
extracted from Fig.~2(b) of \citet{pradhan2015deposition}
using the procedure described above.
Specifically,
Fig.~\ref{fig:comp}(a) compares the density of the deposit
predicted by (\ref{eq:D}) (shown with the black solid curve) with
the experimental results for
the left-hand droplet (shown with the blue dashed curve) and
the right-hand droplet (shown with the red dotted curve).
Given the approximations made
in both the mathematical model and in the processing of the experimental image,
the agreement between theory and experiment
shown in Fig.~\ref{fig:comp}(a) is remarkably good,
especially when it is noted that {\it no} fitting parameters have been used.
The only scaling used is to ensure that the total mass
(not given by \citet{pradhan2015deposition}) is the same for theory and experiment.

While the agreement between theory and experiment
shown in Fig.~\ref{fig:comp}(a) is already good,
with an integral absolute relative error of around
7\% for the left-land droplet and
9\% for the right-hand droplet,
the experimental results are inevitably rather noisy.
Figure~\ref{fig:comp}(b) compares the density of the deposit
predicted by (\ref{eq:D}) (shown again with the black solid curve) with
the experimental results averaged across both droplets,
as well as about $\phi=0$ (shown with the black dashed curve).
In particular,
Fig.~\ref{fig:comp}(b) shows that, as expected,
averaging reduces the noise in the experimental results and,
rather pleasingly,
leads to even better agreement between theory and experiment
than that shown in Fig.~\ref{fig:comp}(a),
with an integral absolute relative error of around 3\%.
Figure~\ref{fig:comp}(b) also includes
the corresponding phenomenological estimate of the radially-integrated fluid flux
(now interpreted as the density of the deposit)
derived by \citet{wray2020competitive}
by using the approach of \citet{saenz2017dynamics},
denoted here by $R_1(\phi,b)$, given by
\begin{equation}
R_1(\phi,b) = \int_0^1 \calJ(r)
\left[1-\frac{F\sqrt{b^2-1}}{2\pi\left(\rho^2+b^2-2br\cos\phi\right)}\right]
r \, \rd{r},
\end{equation}
which can be evaluated to yield
\begin{equation}\label{eq:estimate}
R_1(\phi,b) = \frac{2}{\pi} \left[
1 - \frac{F\sqrt{1-k^2}}{\pi^2\sin\phi} \mathrm{Im}
\left\{\frac{\log\left[-\left(k\ee^{-i\phi}
+ \sqrt{k^2\ee^{-2i\phi}-1}\right)\right]}
{\sqrt{k^2\ee^{-2i\phi}-1}}\right\} \right],
\end{equation}
where
$k = 1/b \, (< 1)$ and
$F$ is given by (\ref{eq:integralTwoDropFlux}).
In particular,
Fig.~\ref{fig:comp}(b) shows that
the estimate (\ref{eq:estimate}) is reasonably accurate,
with an integral absolute relative error of around 9\%,
and captures the experimental results qualitatively but not
quite quantitatively.

\section{A triplet of identical droplets}
\label{sec:triplet}

\begin{figure}[tp]
\begin{center}
\includegraphics[width=0.85\textwidth]{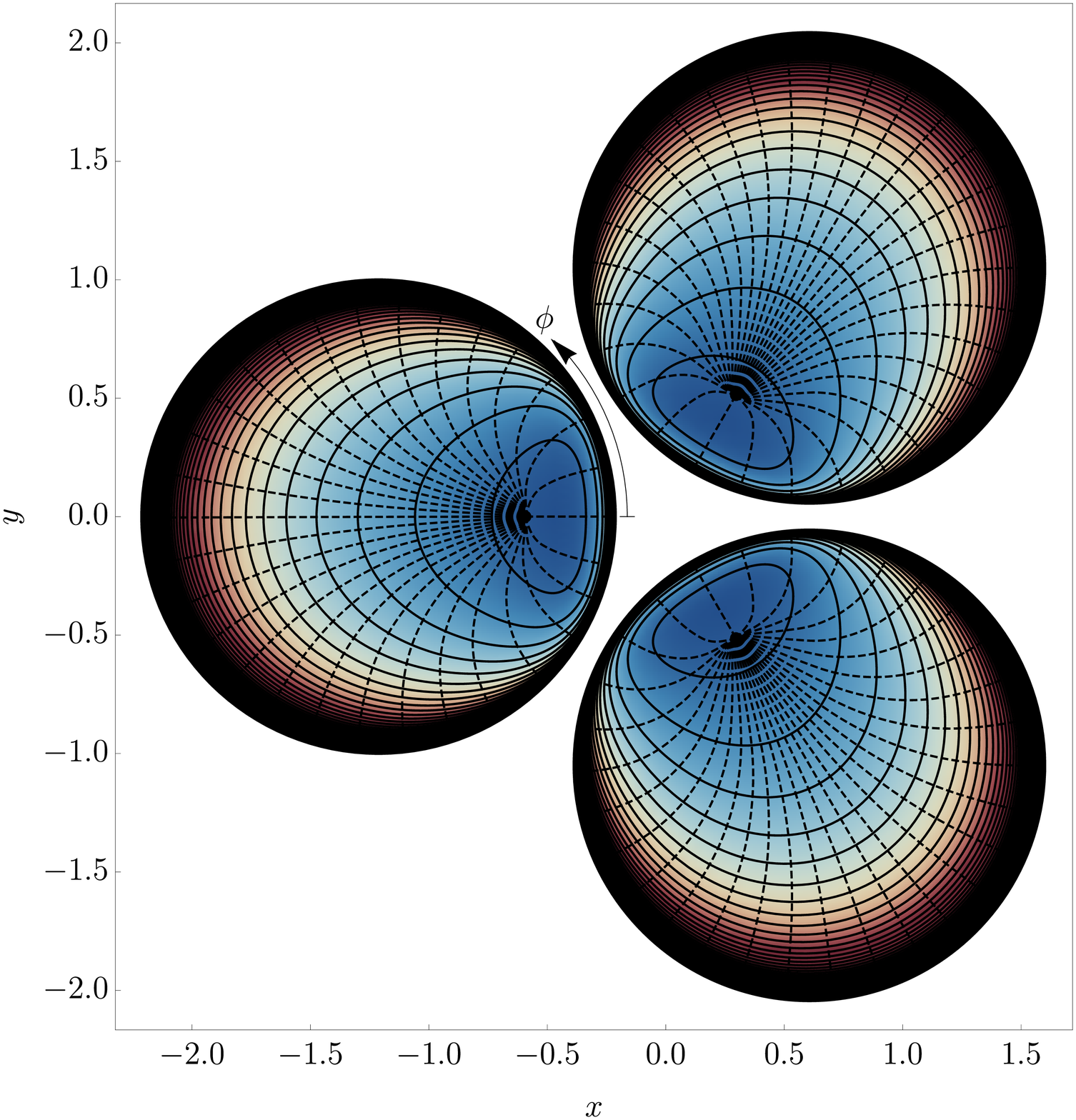} \hspace{0.25cm}
\raisebox{\height}{\includegraphics[width=0.065\textwidth]{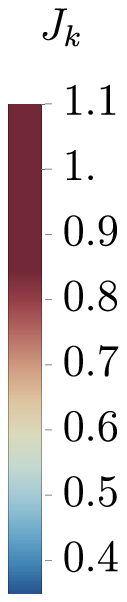}}
\end{center}
\caption{
Contours of the local evaporative fluxes
$J_k$ for $k=1,2,3$ (solid curves), and
the resulting streamlines of the depth-averaged flows (dashed curves)
for a triplet of identical droplets of unit radius
with their centres a distance $b\sin(\pi/3)=2.1$ apart.
}
\label{fig:tripletfluxesandstreamlines}
\end{figure}

\begin{figure}[tp]
\begin{center}
(a)
\includegraphics[width=0.9\textwidth]{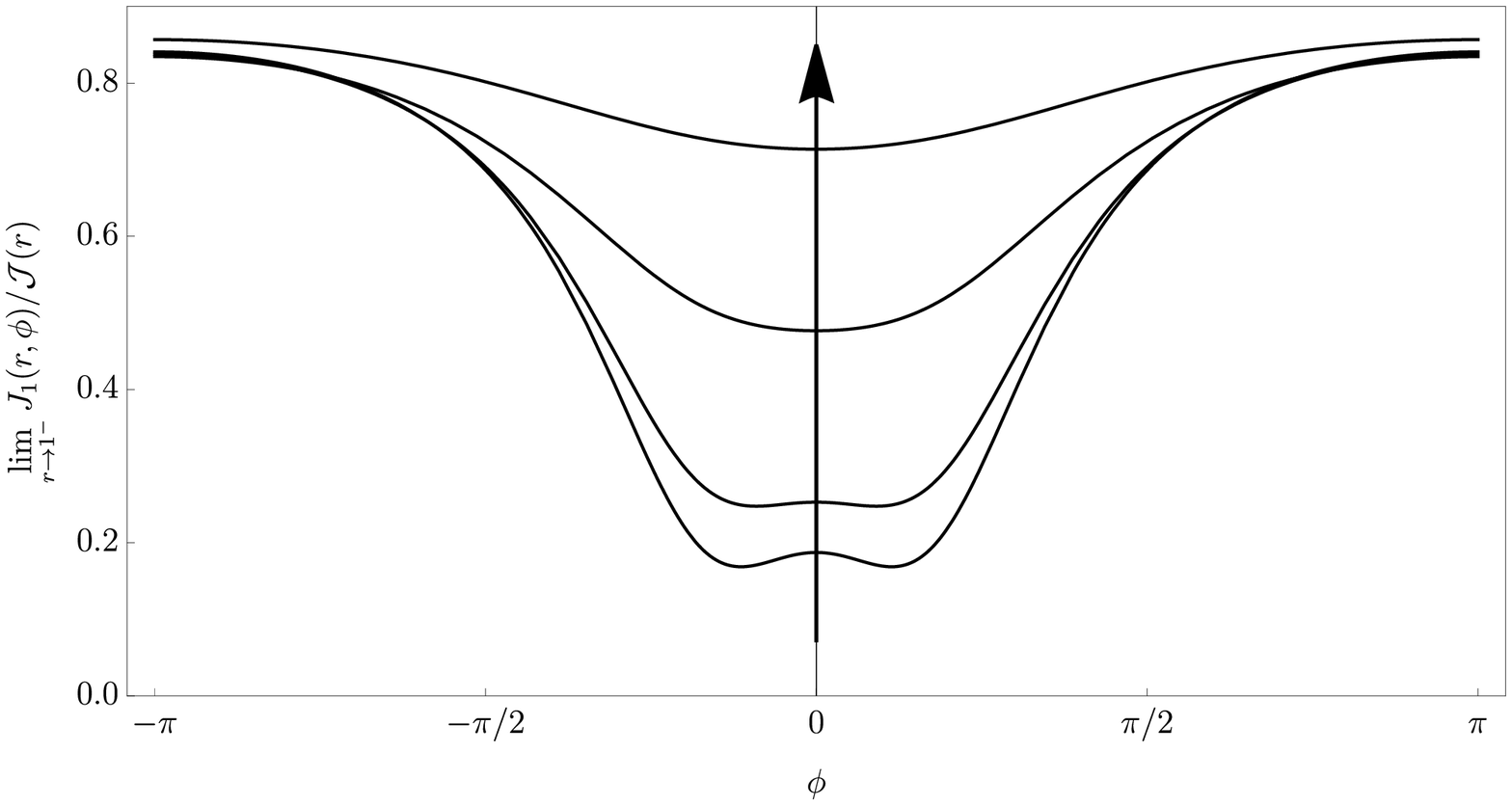}\\
(b)
\includegraphics[width=0.9\textwidth]{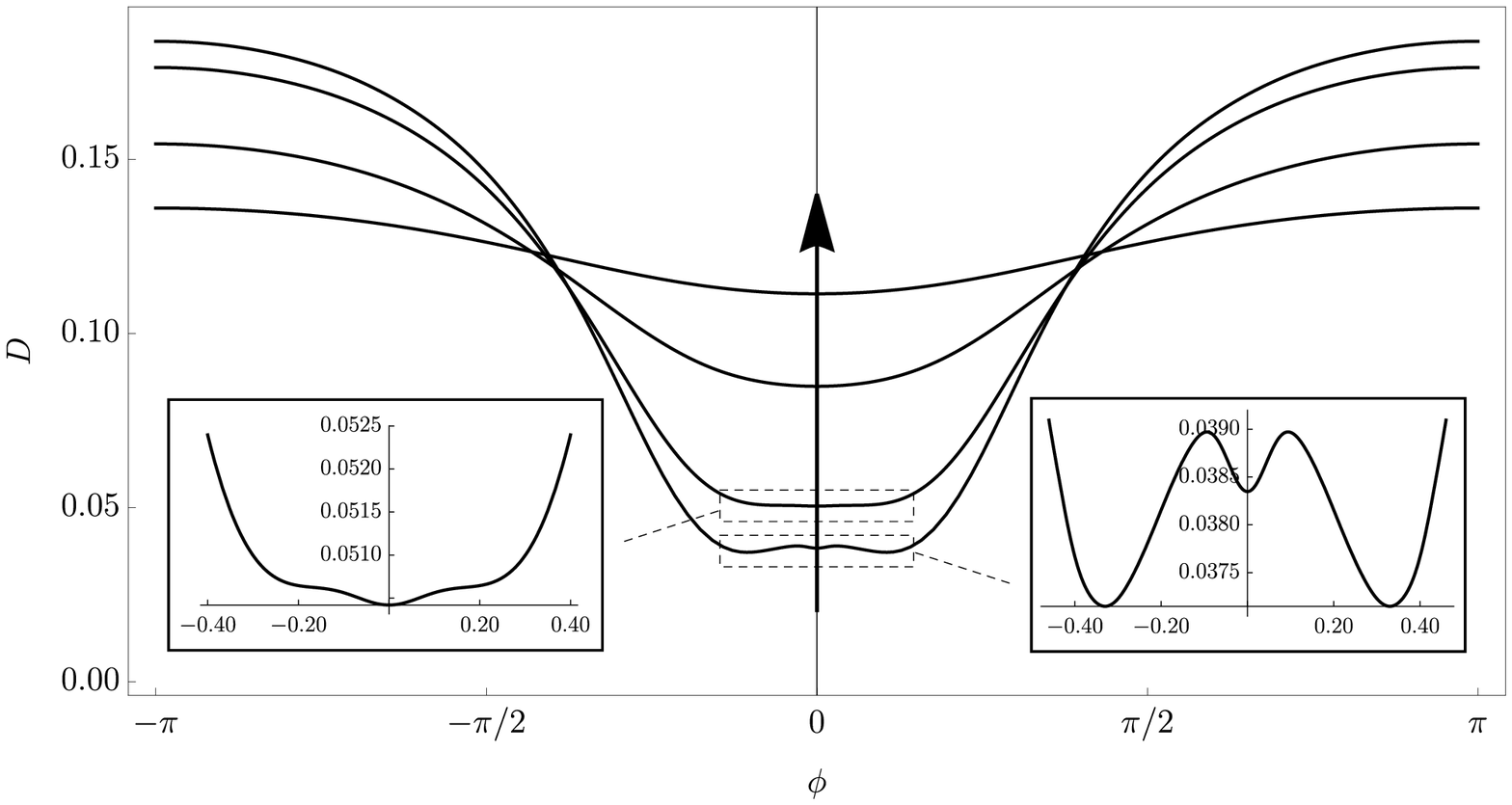}\\
\end{center}
\caption{
(a) The normalised evaporative flux at the contact line,
$\lim_{r \to 1^-}J_1(r,\phi)/\calJ(r)$,
and
(b) the density of the deposit $D$
as functions of the azimuthal coordinate $\phi$
for a triplet of identical droplets of unit radius
with their centres distances $b\sin(\pi/3) = 2.1$, $2.25$, $3$ and $5$ apart.
In both parts the arrow indicates the direction of increasing $b$.
The insets show enlargements of $D$ near $\phi=0$ when $b\sin(\pi/3)=2.1$ and $2.25$.
}
\label{fig:tripletfluxanddensityofdeposit}
\end{figure}

In this Section we use the same approach as that described
in Sec.~\ref{sec:pair}
to determine the densities of the deposits from a
triplet of identical droplets of unit radius with their centres
located at $(-b/2,0)$ and $(b/2)(\cos(\pi/3),\pm\sin(\pi/3))$,
\ie with their centres a distance $r_{1,2}=r_{2,3}=r_{3,1}=b\sin(\pi/3) \, (>2)$ apart.
Note that, in the terminology used by \citet{wray2020competitive},
the centres of the droplets lie at the vertices of an equilateral triangle
with side $b\sin(\pi/3)$ and circumradius $b/2$.
Similarly to in Sec.\ \ref{sec:formulation},
by symmetry it is sufficient to consider only the left-most droplet
with its centre located at $(-b/2,0)$,
corresponding to $k=1$.
The results are summarised in
Figs.~\ref{fig:tripletfluxesandstreamlines}
and
\ref{fig:tripletfluxanddensityofdeposit},
which show
contours of
the local evaporative fluxes $J_k$ for $k=1,2,3$ and
the resulting streamlines of the depth-averaged flows
for $b\sin(\pi/3)=2.1$,
and
the normalised evaporative flux at the contact line,
$\lim_{r \to 1^-}J_1(r,\phi)/\calJ(r)$,
and
the density of the deposit $D$
as functions of the azimuthal coordinate $\phi$
for several values of $b$,
respectively.
In particular,
Figs.~\ref{fig:tripletfluxesandstreamlines}
and
\ref{fig:tripletfluxanddensityofdeposit}
show that, as expected,
the shielding effect again reduces the local evaporative flux and
leads to a spatially non-uniform deposit at the contact line.
However,
in this case the effect is evidently more subtle than that
described in Sec.~\ref{sec:pair} for a pair of droplets.
Specifically,
when the droplets are sufficiently far apart,
the shielding effect reduces the local evaporative flux
the most in the direction towards the centre of the triangle
(\ie at $\phi=0$)
and
the least in the direction away from the centre of the triangle
(\ie at $\phi=\pi$),
but these directions do {\it not} now correspond
to those in which the droplets are closest together
(\ie at $\phi=\pm\pi/6$).
However,
when the droplets are sufficiently close together,
the local evaporative flux develops two (symmetric) local minima
at non-zero values of $\phi$
(as shown by the curve for $b\sin(\pi/3)=2.1$
in Fig.~\ref{fig:tripletfluxanddensityofdeposit}(a)), and
the density of the deposit develops local maxima and mimima
(as shown by the curve for $b\sin(\pi/3)=2.1$
in Fig.~\ref{fig:tripletfluxanddensityofdeposit}(b)).
Note that, as in Fig.~\ref{fig:pairfluxanddensityofdeposit}(b),
by conservation of mass, the total mass of the deposit
is the same for all of the values of $b$ used in Fig.~\ref{fig:tripletfluxanddensityofdeposit}(b).

\section{Conclusions}
\label{sec:conclusions}

In the present work
we obtained theoretical predictions for the spatially non-uniform
densities of the contact-line deposits left on the substrate
after the competitive diffusion-limited evaporation of
multiple thin axisymmetric sessile droplets in proximity to each other.
In particular,
we gave predictions for the deposits
from a pair of identical droplets, which showed
that the deposit is reduced the most where the droplets are closest together, and
demonstrated excellent quantitative agreement
with experimental results of \citet{pradhan2015deposition}.
We also gave corresponding predictions for a triplet of identical
droplets arranged in an equilateral triangle, which showed
that the effect of shielding on the deposit
is more subtle in this case.

We note that, while the present analysis is formally restricted to thin droplets,
the fact that much of the deposition from a non-thin droplet occurs towards the end of its lifetime
when the contact angle is small and the velocity within the droplet is large
(sometime referred to as ``the rush hour'', see, for example, \citep{hamamoto2011velocity,marin2011rushhour})
means that the results of the present analysis
are also expected to provide useful predictions for the contact-line deposits from non-thin droplets.

A brief observation about the validity of the diffusion-limited model
of evaporation is perhaps appropriate here.
The excellent agreement with experimental results of \citet{khilifi2019study}
found by \citet{wray2020competitive} attests to the accuracy
of the theoretical predictions for the {\it integral} evaporative flux,
but it does not tell us anything directly about the accuracy of the
theoretical predictions for the spatial distribution of the
{\it local} evaporative flux or the resulting fluid flow within the droplet.
For an isolated axisymmetric droplet,
the fluid flow within the droplet is axisymmetric and
the density of the deposit depends only on
the initial distribution of particles and so, in particular,
the density of the deposit does not depend on the details of the fluid flow.
However,
in non-axisymmetric situations,
such as the non-axisymmetric droplets considered by \citet{saenz2017dynamics}
and
the non-axisymmetric evaporation of multiple droplets
considered in the present work,
understanding the details of the fluid flow
is essential to determining the density of the deposit.
Thus the comparison with experimental results described in the present work
is perhaps the most stringent test of the diffusion-limited model to date,
a test which it evidently passes remarkably well.

Finally,
we note that the approach described in the present work is rather general
and can, in principle, be applied to any arrangement of any number of
thin droplets with pinned or unpinned contact lines.

\section*{Acknowledgements}

The authors gratefully acknowledge valuable discussions with
Hannah-May D'Ambrosio (University of Strathclyde)
and
Prof.\ Khellil Sefiane (University of Edinburgh)
about various aspects of droplet evaporation.

\begin{appendix}

\section{Solution in the limit of large Bond number}
\label{sec:applargeBo}

\renewcommand{\theequation}{A.\arabic{equation}}

\setcounter{table}{-1}

As mentioned at the end of Sec.~\ref{sec:introduction},
the present analysis is for the most commonly studied case of small droplets
in which capillary effects dominate over gravitational effects,
corresponding to the limit of small Bond number.
In this Appendix we describe
the corresponding analysis in the case of large droplets,
for which gravitational effects dominate over capillary effects,
corresponding to the limit of large Bond number,
in which even greater analytical progress is possible.

In the limit of large Bond number,
the free surface of the droplet is flat,
\ie $h_k = h_k(t)$,
except in a narrow region near the contact line
which we may neglect \citep{rienstra1990shape}, and so
the volume of the (nearly cylindrical) droplet is now given by
$V_k = \pi a_k^2 h_k$.
The local fluid fluxes are again given by (\ref{eq:fluxExps1}),
while the kinematic condition (\ref{eq:kinPartial}) simplifies to
\begin{equation}
\dd{h_k}{t} - \frac{h_k^3}{3}\nabla^2 p_k = -J_k,
\end{equation}
and hence (\ref{eq:dhdtequation}) becomes
\begin{equation}\label{eq:dhdkequationlarge}
\dd{h_k}{t} = -\frac{1}{\pi a_k^2}
\int_{\phi=0}^{\phi=2\pi} \int_{r=0}^{r=a_k} r \, J_k
\, \rd{r} \, \rd{\phi},
\end{equation}
and so the partial differential equation for $p_k$
given by (\ref{eq:pkequation}) becomes
\begin{equation}\label{eq:pkequationlarge}
\nabla^2 p_k = \frac{3}{h_k^3}\left[J_k - \frac{1}{\pi a_k^2}
\int_{\phi=0}^{\phi=2\pi} \int_{r=0}^{r=a_k} r \, J_k
\, \rd{r} \, \rd{\phi} \right].
\end{equation}

In principle,
the same approach as that used in the main body of the present work
can be used to solve the corresponding problem for large droplets.
However, for brevity, in this Appendix we simply show how to obtain
explicit asymptotic expressions for the pressure, and hence for
the fluid fluxes and the density of the deposit,
for a well-separated pair of identical droplets.

Adopting the same notation as in Sec.~\ref{sec:pair},
and, without loss of generality, taking $a_1=a_2=1$ and $h_1(0)=h_2(0)=1$,
the evaporative flux from the left-hand droplet
is again given by equation (\ref{eq:pairJ1}),
which can be expanded as
\begin{equation}\label{eq:J1pairlarge}
J_1 = \frac{2}{\pi\sqrt{1-r^2}} \left[ 1 - \frac{2}{\pi b}
+ \frac{4}{\pi^2 b^2}\left(1 - \pi r \cos \phi\right)\right]
+ O\left(b^{-3}\right)
\end{equation}
in the limit of well-separated droplets, $b \to \infty$.
Hence, the equation for the pressure in the left-hand droplet,
obtained by setting $k=1$ in (\ref{eq:pkequationlarge}), can be expanded as
\begin{align}
\nabla^2 p_1 = \frac{6}{\pi h^3}\left[\frac{1}{\sqrt{1-r^2}}
\left( 1 - \frac{2}{\pi b} + \frac{4}{\pi^2 b^2}\left(1-\pi r\cos\phi\right) \right)
- 2\left( 1 - \frac{2}{\pi b} + \frac{4}{\pi^2 b^2} \right)\right]
+ O\left(b^{-3}\right),
\end{align}
with solution
\begin{align}
\frac{\pi h^3}{3}p_1 = &\left( 1 - \frac{2}{\pi b} + \frac{4}{\pi^2 b^2} \right)
\left[2\log\left(1 + \sqrt{1-r^2}\right) - r^2 - 2\sqrt{1-r^2}\right]
\nonumber\\
& \mbox{}
+ \frac{8}{3\pi r b^2}
\left[1 + r^2 - \left(1-r^2\right)^{3/2}\right]\cos\phi
+ O\left(b^{-3}\right).
\label{eq:pAnaSol}
\end{align}
Hence the local fluid fluxes can be expanded as
\begin{align}
Q_1^{(r)} &= 2\left(1 - \frac{2}{\pi b} + \frac{4}{\pi^2 b^2}\right)
\frac{- 1 + r^2 + \sqrt{1-r^2}}{\pi r}
\nonumber\\
& \mbox{} + \frac{8}{3\pi^2 r^2b^2}
\left[1 - r^2 - \left(1+2r^2\right)\sqrt{1-r^2}\right]\cos\phi
+ O\left(b^{-3}\right)
\end{align}
and
\begin{align}
Q_1^{(\phi)} &= \frac{8}{3\pi^2 r^2b^2}
\left[1 + r^2 - \left(1-r^2\right)^{3/2}\right]\sin\phi
+ O\left(b^{-3}\right).
\end{align}

We now seek to determine the density of the deposit
at a point on the contact line with polar angle $\phi=\phiC$, \ie $D(\phiC)$,
where $D$ is given by (\ref{eq:D}) with $M$ given by (\ref{eq:M}).
Note that $M(2\pi)=\pi$ for this cylindrical droplet.
In order to do this we must first locate
the source from which all of the streamlines emanate.
By symmetry this must lie on the line of symmetry, $\phi=0$, and,
by expanding $Q_1^{(r)} = 0$ in powers of $b$,
we find that it is located at $r=\rS$ and $\phi=\phiS=0$, where
\begin{equation}
\rS = \frac{20}{3\pi b^2}
+ O\left(b^{-3}\right).
\end{equation}
The streamline $\phi = \phiS(r;\phiC)$
starting at the source at $r=\rS$ and $\phi=\phiS=0$ and
ending on the contact line at $r=1$ and $\phi=\phiS=\phiC$
satisfies
\begin{equation}
\dd{\phiS}{r} = \frac{1}{r}\frac{Q^{(\phi)}}{Q^{(r)}}
= \frac{4\left(3 - r^2 + 2\sqrt{1-r^2}\right)}
{3\pi r^2 \sqrt{1-r^2} \, b^2}\sin\phiS
+ O\left(b^{-3}\right),
\end{equation}
which may be solved by expanding $\phiS$ in powers of $b$ to obtain
\begin{equation}
\phiS(r;\phiC) = \phiC
+ \frac{4\left[r\left(2 + \arccos r\right) - 2 - 3\sqrt{1-r^2}\right]}
{3\pi r b^2}\sin\phiC
+ O\left(b^{-3}\right).
\end{equation}
The deposit at $\phi=\phiC$ is then given by
\begin{equation}
D(\phiC) = \frac{\rd M}{\rd\phiC}
= \frac{\rd}{\rd\phiC}\int_{r=\rS}^{r=1}\int_{\phi=0}^{\phi=\phiS(r;\phiC)} r
\, \rd{\phi} \, \rd{r},
\end{equation}
and hence
\begin{equation}\label{eq:Dlarge_2}
D(\phiC) = \int_{\rS}^{1} r \, \frac{\partial \phi_S}{\partial \phi_C}
\, \rd{r}
= \frac{1}{2}
- \frac{5\pi+8}{6\pi b^2}\cos\phiC
+ O\left(b^{-3}\right).
\end{equation}
The expansion (\ref{eq:Dlarge_2}) can be continued to next order to obtain
\begin{equation}\label{eq:Dlarge_3}
D(\phiC) = \frac{1}{2}
- \frac{5\pi+8}{6\pi b^2}\cos\phiC
- \frac{4(13+\log 4)}{15\pi b^3}\cos(2\phiC)
+ O\left(b^{-4}\right).
\end{equation}
Higher-order corrections to $D(\phiC)$ may also be obtained,
but rapidly become more complicated.
Table \ref{tab:table} shows a comparison between the coefficients of the terms
$b^{-2}\cos\phiC$ and
$b^{-3}\cos(2\phiC)$
obtained from
the analytical prediction for $D$ given by (\ref{eq:Dlarge_3})
and
the values of $D$ obtained from (\ref{eq:M}) and (\ref{eq:D})
for $b=3$ and $b=10$.
The diminishing discrepancies as $b$ increases evident in Table \ref{tab:table}
are due to the omitted higher-order corrections.
In particular,
analysis of these corrections indicates that
the expression for $b^{-2}\cos\phiC$
has relative error $O\left(b^{-2}\right)$, while
the expression for $b^{-3}\cos(2\phiC)$
has relative error $O\left(b^{-1}\right)$,
explaining the superior performance of the former.

\renewcommand{\arraystretch}{1.8}
\begin{table}[t]
\begin{longtable}{|c|c|c|c|} \hline
XXXXTermXXXX &
XXXXAnalytical (AXX)XXXX &
XXXXXXXX &
XXXXXXXX \kill
Term &
Analytical (\ref{eq:Dlarge_3}) &
$b=3$ &
$b=10$ \\ \hline
$b^{-2}\cos\phiC$ &
$\displaystyle-\frac{5\pi+8}{6\pi}\simeq-1.25775$
& $-1.28$ & $-1.26$ \\ \hline
$b^{-3}\cos(2\phiC)$ &
$\displaystyle-\frac{4(13+\log 4)}{15\pi}\simeq-1.22115$ &
$-1.06$ &  $-1.13$ \\ \hline
\end{longtable}
\caption{
Comparison between the coefficients of the terms
$b^{-2}\cos\phiC$ and
$b^{-3}\cos(2\phiC)$
obtained from
the analytical prediction for $D$ given by (\ref{eq:Dlarge_3})
and
the values of $D$ obtained from (\ref{eq:M}) and (\ref{eq:D})
for $b=3$ and $b=10$.
}
\label{tab:table}
\end{table}

\end{appendix}

\bibliography{multipleCoffee}

\end{document}